%% file: main.tex
\documentclass[sigconf]{acmart}

\copyrightyear{2024}
\acmYear{2024}
\setcopyright{rightsretained}
\acmConference[CCS '24] {Proceedings of the 2024 ACM SIGSAC Conference on Computer and Communications Security}{October 14--18, 2024}{Salt Lake City, UT, USA.}
\acmBooktitle{Proceedings of the 2024 ACM SIGSAC Conference on Computer and Communications Security (CCS '24), October 14--18, 2024, Salt Lake City, UT, USA}
\acmISBN{979-8-4007-0636-3/24/10}
\acmDOI{10.1145/3658644.3670285}

\usepackage{url}
\usepackage{threeparttable}
\usepackage{multirow}
\usepackage{soul}
\usepackage{enumitem}
\usepackage{xspace}
\usepackage{tcolorbox}
\usepackage{pifont}
\usepackage{subcaption} 

\usepackage{colortbl}
\definecolor{lightgray}{gray}{0.9}
\definecolor{white}{gray}{1}
\usepackage{cleveref}

\definecolor{mycolor}{RGB}{241, 242, 243}
\sethlcolor{mycolor}

\newcommand{\sys}{{\texttt SafeEar}\xspace}

\AtBeginDocument{%
  \providecommand\BibTeX{{%
    \normalfont B\kern-0.5em{\scshape i\kern-0.25em b}\kern-0.8em\TeX}}}

\newenvironment{icompact}{
  \begin{list}{$\bullet$}{
    \itemindent 0em
    \itemsep 3pt
    \leftmargin 0.15in}
      }
{\normalsize
\end{list}
}

\newcommand{\blue}[1]{\textcolor{black}{#1}}

\begin{document}

\title{\sys: Content Privacy-Preserving Audio Deepfake Detection}


\author{Xinfeng Li}
\authornote{Equal Contribution.}
\affiliation{%
  \institution{Zhejiang University}
  \city{HangZhou}
  \state{Zhejiang}
  \country{China}
  \postcode{310058}
}
\email{xinfengli@zju.edu.cn}

\author{Kai Li}
\authornotemark[1]
\affiliation{
  \institution{Tsinghua University}
  \city{Beijing}
  \country{China}
}
\email{tsinghua.kaili@gmail.com}

\author{Yifan Zheng}
\affiliation{
  \institution{Zhejiang University}
  \city{HangZhou}
  \state{Zhejiang}
  \country{China}
  \postcode{310058}
}
\email{zhengyf@zju.edu.cn}

\author{Chen Yan}
\authornote{Chen Yan is the Corresponding Author.}
\affiliation{
  \institution{Zhejiang University}
  \city{HangZhou}
  \state{Zhejiang}
  \country{China}
  \postcode{310058}
}
\email{yanchen@zju.edu.cn}

\author{Xiaoyu Ji}
\affiliation{%
  \institution{Zhejiang University}
  \city{HangZhou}
  \state{Zhejiang}
  \country{China}
  \postcode{310058}
}
\email{xji@zju.edu.cn}

\author{Wenyuan Xu}
\affiliation{%
  \institution{Zhejiang University}
  \city{HangZhou}
  \state{Zhejiang}
  \country{China}
  \postcode{310058}
}
\email{wyxu@zju.edu.cn}


\input{sections/abstract}





\maketitle


\input{sections/introduction}
\input{sections/background}

\input{sections/threatmodel}
\input{sections/preliminary_analysis}
\input{sections/design_v2}
\input{sections/exp_setup}
\input{sections/evaluation}
\input{sections/discussion}
\input{sections/related_work}
\input{sections/conclusion}
\bibliographystyle{ACM-Reference-Format}
\bibliography{reference/reference}
\input{sections/appendix}

\end{document}

%% file: sections/abstract.tex
\begin{abstract}
Text-to-Speech (TTS) and Voice Conversion (VC) models have exhibited remarkable performance in generating realistic and natural audio. However, their dark side, audio deepfake poses a significant threat to both society and individuals.
Existing countermeasures largely focus on determining the genuineness of speech based on complete original audio recordings, which however often contain private content. This oversight may refrain deepfake detection from many applications, particularly in scenarios involving sensitive information like business secrets.
In this paper, we propose \sys, a novel 
framework that aims to detect deepfake audios without relying on accessing the speech content within. 
Our key idea is to devise a neural audio codec into a novel decoupling model that well separates the semantic and acoustic information from audio samples, and only use the acoustic information (\textit{e.g.}, prosody and timbre) for deepfake detection. In this way, no semantic content will be exposed to the detector. 
\blue{To overcome the challenge of identifying diverse deepfake audio without semantic clues, we enhance our deepfake detector with real-world codec augmentation.} 
Extensive experiments conducted on four benchmark datasets demonstrate \sys's effectiveness in detecting various deepfake techniques with an equal error rate (EER) down to 2.02\%. 
Simultaneously, it shields five-language speech content from being deciphered by both machine and human auditory analysis, demonstrated by word error rates (WERs) all above 93.93\% and our user study.
Furthermore, our benchmark constructed for anti-deepfake and anti-content recovery evaluation helps provide a basis for future research in the realms of audio privacy preservation and deepfake detection.
\end{abstract}

%% file: sections/introduction.tex

\section{Introduction}
Recent advances in text-to-speech (TTS) and voice conversion (VC) technologies have enabled the generation of highly realistic and natural-sounding speech, imitating specific individuals saying things they never actually said. However, such technologies have been misused to create audio deepfakes, posing significant security threats. 
For instance, deepfakes disseminated on the Internet can manipulate public opinion, serving purposes like propaganda, defamation, or terrorism~\cite{suwajanakorn2017synthesizing, PoliticDeepFake}.	
Besides, audio deepfake fraud in calls and virtual meetings, including a notable UK case where \$35 million was stolen using a cloned CEO's voice~\cite{CEOscam}, has financially affected 7. 7\% individuals, according to a 2023 McAfee survey~\cite{McAfeeReport}. 
These have spurred the development of diverse audio deepfake detection models, designed to discern synthetic from genuine voices and promptly alert potential victims. However, existing works~\cite{AASIST,RawNet2,Rawformer,LCNN-LSTM,ASVspoof2021-baselines} typically take audio waveforms or spectral features (\textit{e.g.}, LFCC~\cite{LFCC_SE-ResNet34}) as inputs, which require accessing complete speech information. These approaches, while efficient, \blue{raise substantial privacy concerns due to the potential exposure of private speech content, particularly in virtual communications that involve user privacy like business secrets or medical conditions~\cite{Amazon_leak}}. Thus, despite current detectors' utility in thwarting deepfakes, there is natural hesitancy in using them due to the risk of content leakage.

\begin{figure}[t]
    \centering
    \includegraphics[width=0.42\textwidth]{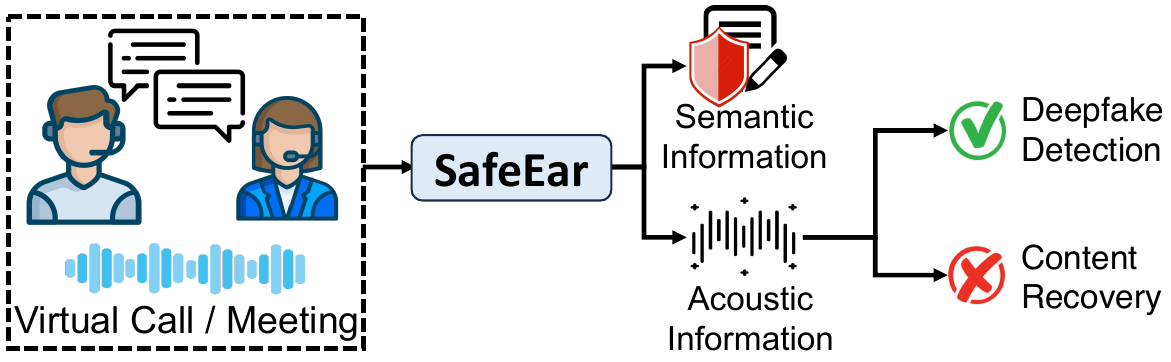}
    \caption{\sys framework decouples speech samples into semantic and acoustic information. By using acoustic-only information, \sys achieves reliable deepfake detection while protecting user content privacy from recovery attacks.}
    \label{fig:fig1}
    \vspace{-15pt}
\end{figure}

In this paper, we introduce \sys\footnote{Our demo, code, and dataset are available on \url{https://SafeEarWeb.github.io/Project/}.}, a novel framework designed to effectively detect audio deepfakes while preserving content privacy. As shown in Figure~\ref{fig:fig1}, the key idea of \sys is to decouple speech into semantic and acoustic information. This approach enables reliable deepfake detection using processed acoustic information while preventing potential adversaries from accessing the semantic content, even if they employ advanced automatic speech recognition (ASR) models or human auditory analysis. \blue{Thus, \sys is particularly suited for third-party audio service scenarios where an honest-but-curious server might offer reliable deepfake detection service, yet unethically eavesdrops user speech content. For detection services operated on trusted local devices, the \sys framework also provides an extra layer of protection for user privacy.}


To our knowledge, this is the first work to develop a content privacy-preserving audio deepfake detection framework. \sys is inspired by the intuition that audio deepfakes aim to replicate a speaker's timbre and prosody disregarding the speech content. In contrast, speech recognition systems focus on extracting semantic content, independent of the speaker-related features. This dichotomy indicates that these two tasks may rely on mutually independent features, suggesting the potential for designing an effective audio deepfake detector analyzing only acoustic information without exposing semantic content. 
However, materializing \sys is challenging in two aspects.



\textit{How to protect content privacy from recovery by adversaries?}
\sys aims to safeguard speech content privacy against both machine-based and human auditory analysis. Prior works using adversarial examples~\cite{carlini2018audio,li2024vrifle,zheng2023silent} for ASR model disruption have shown limited effectiveness against human listeners. \sys tackles this by decoupling speech into semantic and acoustic tokens and provides only acoustic tokens to the detector, where tokens mean the discrete representations of information~\cite{van2017neural}. Consequently, although content recovery adversaries can receive a series of acoustic tokens, the lack of semantic clues hinder their recovery of understandable content. This approach, along with randomly shuffling the acoustic tokens, further obfuscates the contextual patterns that both machine-based and human auditory analysis rely on for content comprehension~\cite{Nature_contextual}. \sys also defends against a range of adversaries who might use decoders to transform acoustic tokens into speech waveforms and analyze them.

\textit{How to deliver accurate deepfake detection merely based on acoustic tokens?}
The challenge lies in the absence of semantic information and the disrupted acoustic patterns (\textit{e.g.}, timbre and prosody) due to shuffling. These content protection strategies may complicate the identification of clues necessary to differentiate genuine from synthetic audio. 
\blue{We address this by developing a Transformer-based detector and identifying its optimal number of multi-head self-attention (MHSA)~\cite{vaswani2017attention} for processing acoustic-only inputs. This adaptation allows the deepfake detector to better capture dynamic spatial weighting and local-global feature interactions.}
Additionally, deepfakes can occur across various communication platforms, which can degrade the deepfake-and-genuine gap due to the effects of codec compression like G.722~\cite{G.722} and OPUS~\cite{Opus} during audio transmission.
To address this, we strategically integrate several representative codecs into our training pipeline to counteract the disruptive effects of codecs, ensuring \sys's accuracy and reliability across diverse real-world scenarios. 

We construct a comprehensive benchmark to compare the performance of \sys and other systems in deepfake detection and content privacy protection. This benchmark comprises four datasets, including three standard datasets---ASVspoof 2019~\cite{ASVspoof2019_dataset}, ASVspoof 2021~\cite{ASVspoof2021_dataset} for deepfake detection, Librispeech~\cite{librispeech} for content protection, and CVoiceFake we established for both aspects. CVoiceFake is a multilingual deepfake dataset sourced from the CommonVoice dataset~\cite{Commonvoice} with over 1.25 million bonafide and deepfake voice samples in five languages. CVoiceFake also includes ground-truth textual transcriptions, making it also an ideal benchmark against content recovery attacks. To our knowledge, CVoiceFake fills the gap in cross-language deepfake datasets~\cite{ADD_survey}, and we hope it can serve as a basis to assist future research in this area.

Based on the above benchmark data\-sets, our extensive experiments focus on two critical tasks: deepfake detection and content protection. 
For the deepfake detection task, we benchmark \sys against eight baseline detectors across three deepfake datasets, which feature a variety of deepfake speech samples generated using popular TTS and VC technologies. Specifically,  \sys achieves comparable performance with top-tier deepfake detectors based solely on acoustic information, with an optimal equal error rate (EER) as low as 2.02\%. 
Regarding the content protection task, we evaluate \sys's efficacy against three levels of content recovery adversaries: \textit{naive} (CRA1), \textit{knowledgeable} (CRA2), and \textit{adaptive} (CRA3), thwarting all content recovery attempts with word error rates (WERs) above 93.93\%. \sys also demonstrates robustness in safeguarding speech content in English and four extra unseen languages, suggesting its potential for wider application. The benchmark and experiment audio samples can be found on our demo website~\cite{safeear_demo}. 

\textbf{Summary of Contributions.} Our technical and experimental contributions are as follows:
\begin{icompact}
    \item To our knowledge, we make the first attempt to investigate and validate the feasibility of achieving audio deepfake detection while preserving speech content privacy. 

    
    \item We propose \sys, a novel privacy-preserving deepfake detection framework that devises a neural audio codec into a semantic-acoustic information decoupling model, ensuring content privacy. We further develop an advanced detector that achieves effective deepfake detection with only acoustic information.

    \item We construct CVoiceFake and establish a comprehensive benchmark focusing on the deepfake detection and content privacy preservation tasks. Our experiments demonstrate the effectiveness of \sys in detecting deepfake audio under various impact factors and in thwarting multiple content recovery attacks.
    
    

\end{icompact}

%% file: sections/background.tex
\vspace{-10pt}
\section{Background}
\subsection{Audio Deepfake Generation}
Deepfake audios are generated using either text-to-speech (TTS) or voice conversion (VC), where the deployment of deep neural networks (DNN) gradually becomes a dominant method that achieves much better voice quality. 

\textbf{Text-to-Speech:} TTS has a long history and recently advances remarkably due to the evolution of deep learning techniques~\cite{zen2013statistical,fan2014tts,li2019neural}. A typical TTS system can be decomposed into three main components: (1) A frontend text analysis module~\cite{tan2021survey} that converts character into phoneme or linguistic features; (2) An acoustic model~\cite{ren2019fastspeech,li2019neural,xiaoice2} that generates speech features such as Mel filter banks (FBank) or Mel-frequency cepstrum coefficient (MFCC), from either linguistic features or characters/phonemes; (3) A vocoder model~\cite{WORLD,Griffin-Lim,MelGAN,hifiwavegan} that generates waveform from either linguistic features or acoustic features. Additionally, recent progress such as fully end-to-end models~\cite{ren2020fastspeech,kim2021conditional} that directly convert characters/phonemes into waveform, are able to generate high quality audio even close to the human level.


\textbf{Voice Conversion:} VC aims to change some properties of speech, such as speaker identity, emotion, and accents, while reserving the semantic content~\cite{sisman2020overview}. Unlike TTS, the inputs to the VC system is another audio waveform instead of text. VC systems can be roughly categorized into two types regarding the requirement of training data: (1) parallel training data systems require the speech of the same semantic content to be available from both source and target speakers~\cite{tian2017exemplar}; (2) non-parallel training data systems reduce the difficulty of data collection, as no parallel training data is needed. In this scenario, a trainable module designed for disentangling speaker-related features from speech features~\cite{kaneko2017parallel} is necessary to extract pure semantic information, which can be composed with the identity information of other speakers to realize voice conversion.


\subsection{Audio Deepfake Detection}\label{ssec:background_add}
\begin{figure}[t]
    \centering
    \includegraphics[width=0.48\textwidth]{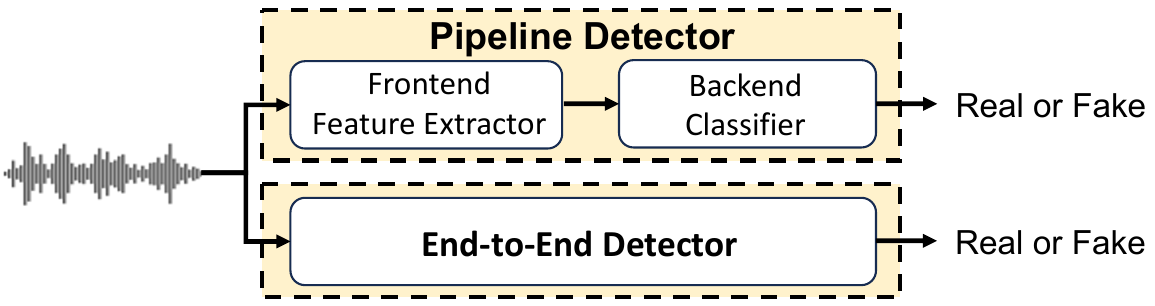}
    \caption{Mainstream solutions on audio deepfake detection: pipeline and end-to-end detector.}
    \label{fig:back_detector}
\end{figure}

Audio deepfake detection is a critical machine learning task that focuses on identifying real utterances from fake ones. An increasing number of attempts~\cite{ADD_survey,AASIST,RawNet2} have been made to further the development of audio deepfake detection. As shown in Figure~\ref{fig:back_detector}, existing mainstream studies on audio deepfake detection can be categorized into two types of solutions: pipeline detector and end-to-end detector. The pipeline solution~\cite{LFCC_SE-ResNet34,LCNN-LSTM,ASVspoof2021-baselines,attention-backend}, consisting of a frontend feature extractor and backend classifier is well established. It extracts spectral features like MFCC and LFCC~\cite{LFCC_SE-ResNet34,LCNN-LSTM}, or token-level Wav2Vec2 features~\cite{xie2021siamese}. In recent years, end-to-end approaches~\cite{AASIST,RawNet2,joint} have attracted more and more attention, which integrates the feature extraction and classification into a single model. This unified approach optimizes the model using raw audio waveforms alongside corresponding real-or-fake labels. 
\sys lies in the pipeline detector group, which fills a gap in privacy-preserving deepfake detection methods. 

\subsection{Speech Representation Decoupling}\label{ssec:background_decouple}
Speech information can be roughly decomposed into three components: content, speaker, and prosody~\cite{liu2023unifyspeech}. Content is semantic information, which can be expressed using text or phonemes. Speaker and prosody features constitute the acoustic information. The former reflects speaker's characteristics such as timbre and volume, while prosody involves intonation, stress, and rhythm of speech, reflecting how the speaker says the content. 
Prior speech representation disentanglement methods mostly leverage a dual-encoder strategy~\cite{qian2019autovc}, where speech is fed into parallel content and speaker encoders to obtain distinct representations. However, this strategy heavily relies on prior knowledge of given languages and speakers and potentially overlooks certain speech information like prosody, which may result in suboptimal decoupling, potentially leading to content leakage or insufficient detection clues. To tackle this issue, \sys presents a novel neural audio codec-based decoupling model that hierarchically decouples speech into semantic and acoustic tokens. It enables content privacy-preserving deepfake detection solely based on acoustic information. In-depth details of our design are elaborated in \S\ref{sec:design}.




%% file: sections/threatmodel.tex
\section{Threat Model}\label{sec:threat_model}


\blue{In this section, we introduce the application scenarios relevant to the \sys framework, and identify two malicious entities posing threats to users, \textit{i.e.}, the \textit{deepfake adversary} (DA) and the \textit{content recovery adversary} (CRA).}

\subsection{Adversary Models}
\hspace{0.4cm}\blue{\textbf{Application Scenarios.} Third-party audio services have become popular in the market because of their advantages in providing specialized functionalities and flexible usage. However, the privacy concern of sharing raw audio with a third party is one of the primary factors preventing users from fully trusting these services, even if the service provider claims to not collect any data. For example, a deepfake detection service provider could be an honest-but-curious content recovery adversary (CRA), detecting deepfake audio to alert victims timely while unethically eavesdropping on conversation content. }

\blue{The \sys framework is designed to relieve such privacy concerns, especially in using third-party audio services. Its frontend decoupling model can be examined and deployed by an entity that is already trusted in processing the raw audio data (e.g., the user's smartphone). Meanwhile, the backend deepfake detector can be operated by any untrusted entities (\textit{i.e.}, detection service providers). In this way, both the detection service and potential adversaries gain access only to the privacy-preserving acoustic tokens, rather than raw audio or unprotected features, which could be easily exploited to recover speech content.}

\textbf{Deepfake Adversary (DA).} The DA's goal is to generate audio that convincingly impersonates real human speakers (TTS) or mimics individuals familiar to the victim (VC). Employing sophisticated TTS and VC models, the adversary can acquire multiple speech samples from a target, using them for voice cloning or create realistic speech for various roles, such as customer service representatives. Moreover, The DA may engage in fraudulent activities on widely used instant communication platforms globally. This introduces two primary detection challenges: (1) Variations in audio codecs across transmission channels can result in different degrees of compression for genuine and deepfake voices, blurring the distinction between them. (2) Deepfake audio in different languages may present unique detection patterns. \blue{Our work does not consider DAs that create adversarial examples to bypass detectors, as it is typically impractical for adversaries to gain knowledge of proprietary, black-box detection systems.} Extensive experiments on deepfake detection using three benchmark datasets are detailed in \S\ref{sec:eval1}.

\textbf{Content Recovery Adversary (CRA).} The CRA seeks to extract intelligible speech content from the acoustic tokens decoupled and shuffled by \sys. Such an adversary could be an honest-but-curious deepfake detection service provider, with prior knowledge of \sys's algorithm. While adversaries receive only the sequences of discrete acoustic tokens, they are capable of reconstructing this feature sequence into speech waveforms using \sys's decoder. Adversaries may also train state-of-the-art ASR models from scratch, and utilize off-the-shelf commercial or local ASR models, to convert the received acoustic tokens into coherent text, or employ human auditory analysis for content recovery. However, they cannot access semantic tokens as \sys does not provide this data. We conduct a comprehensive evaluation of \sys against three levels of content recovery adversaries, as elaborated in \S\ref{sec:eval2}.

\subsection{Defense Goal}
To address the growing concern of deepfake audio in virtual communications, users require detectors to provide reliable alerts. However, there is a natural hesitancy in using them due to the risk of speech content leakage. \sys aims to alleviate this concern by extracting the content-irrelevant features, which can safeguard user content privacy while being suitable for effective detection. \sys's design shall meet two key requirements:


\textbf{Deepfake Detection}: The deepfake detection model in \sys should be finely tuned to work with content-irrelevant features, guaranteeing reliable and accurate detection of deepfake audio.

\textbf{Content Protection:} Features extracted by \sys should be resistant against content recovery attempts by CRAs, regardless of whether they employ machine-based or human auditory methods.



%% file: sections/preliminary_analysis.tex





%% file: sections/design_v2.tex
\section{Design Details}\label{sec:design}
\subsection{Overview of \sys}\label{ssec:design_overview}
\hspace{0.4cm}\textbf{Key Idea.} We aim to propose a framework that achieves two seemingly contradictory objectives: effective deepfake detection and prevention of any attempts at content recovery. Our key idea is to design a novel frontend feature extractor that can decompose speech information into mutually independent discrete representations, \textit{i.e.}, semantic and acoustic tokens, where only the latter being analyzed by subsequent deepfake detectors. Such acoustic tokens can enable effective deepfake detection, but nullify recovery attempts by both machine and human auditory analysis.

\textbf{Intuition Behind \sys.}
The idea of \sys is rooted in a critical insight: audio deepfake technology primarily concentrates on capturing the unique vocal attributes of a target speaker, such as timbre, loudness, rhythm, and pitch, which constitute acoustic information~\cite{liu2023unifyspeech}. However, this technology typically overlooks the actual speech content. 
In fact, several studies have already confirmed the significance of acoustic features in detecting deepfake audios, \textit{e.g.}, timbre~\cite{chaiwongyen2022contribution}, pitch and loudness~\cite{li2022comparative}. In contrast, the core of speech comprehension, both in humans and as modeled in ASR systems, lies in accurately transcribing the semantic content, irrespective of variations in the speaker's acoustic patterns~\cite{yasmin2023effects}. 
The above understanding leads us to believe that developing a deepfake audio detector merely based on acoustic information is feasible. Acoustic information's devoid of semantic content exploitable by adversaries, inherently preserves content privacy.

\textbf{Challenges.} To realize \sys, we faces two challenges. \textit{Challenge 1}: How to design a novel decoupling module that well extracts and secures acoustic tokens, protecting speech content from recovery by machine and human auditory analysis? \textit{Challenge 2}: How to ensure reliable detection against various real-world deepfake audio, despite relying only on acoustic tokens?
\begin{figure}[t]
    \centering
    \includegraphics[width=0.48\textwidth]{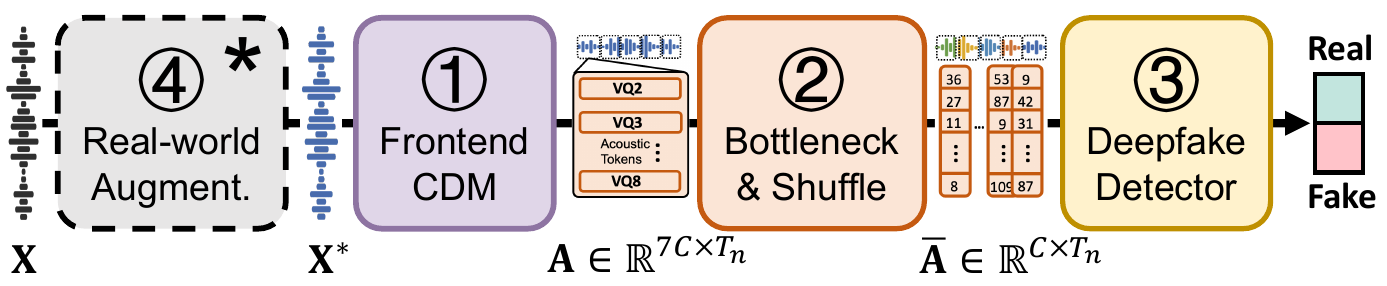}
    \caption{Overview of the \sys framework. In the inference phase, we just need to remove \ding{175}.}
    \label{fig:fig3}
\end{figure}

\textbf{Methodology Outline.} 
As shown in Figure~\ref{fig:fig3}, to address \textit{Challenge 1}, we carefully devise a neural codec architecture (\S\ref{sec:CDM}, \ding{172} in Figure~\ref{fig:fig3}) to flexibly decompose the audio signal $\mathbf{X}\in \mathbb{R}^{1\times T}$ into semantic tokens $\mathbf{S}\in \mathbb{R}^{C\times T_n}$ and acoustic tokens $\mathbf{A}\in \mathbb{R} ^{7C\times T_n}$, where $C$ denotes the token dimension, and $T$ and $T_n$ represent the length of the audio and token, respectively. We combine a bottleneck and shuffle layer (\S\ref{sec:shuffle}, \ding{173} in Figure~\ref{fig:fig3}) to secure the tokens as $\mathbf{\overline{A}}\in \mathbb{R}^{C\times T_n}$, thereby the original content cannot be reconstructed. For \textit{Challenge 2}, we finely tune our backend detector (\S\ref{ssec:design_detection}, \ding{174} in Figure~\ref{fig:fig3}) with optimal number of self-attention heads, as well as mimicking real-world codec transformation from $\mathbf{X}$ to $\mathbf{X^*}$ for the detector training (\S\ref{ssec:codec_augment}, \ding{175} in Figure~\ref{fig:fig3}).


\subsection{Codec-based Decoupling Model (CDM)}\label{sec:CDM}

\begin{figure}[t]
    \centering
    \includegraphics[width=0.45\textwidth]{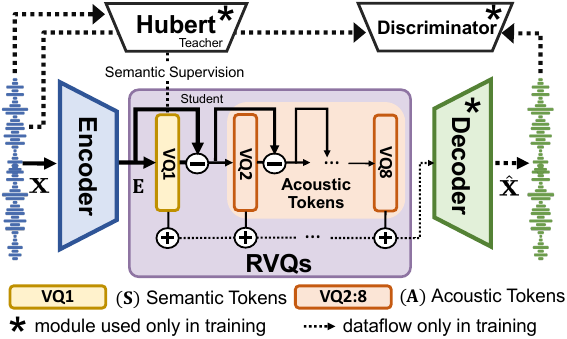}
    \caption{Frontend codec-based decoupling model (\ding{172}) of \sys.}
    \label{fig:CDM}
\end{figure}

Recent advancements in neural audio codecs such as SpeechTokenizer~\cite{zhang2023speechtokenizer}, Encodec~\cite{Encodec} and VALL-E~\cite{wang2023neural} have provided evidence of the advantages of multi-layer residual vector quantizers (RVQs) in accurately representing speech with discrete speech tokens for high-quality and efficient audio transmission, regardless of sound type or language.\footnote{More description of audio codecs are provided in Appendix~\ref{ssec:audio_codec}.}.
We aim to develop the neural codec architecture into an effective decoupling model that separates mixed speech tokens into standalone semantic and acoustic tokens.
\color{black}
As illustrated in Figure~\ref{fig:CDM}, our proposed decoupling model based on the codec architecture (CDM) comprises three core components: an encoder-decoder architecture, a HuBERT-equipped RVQs module, and a discriminator. The encoder-decoder's primary function of precisely reconstructing the original audio compels the encoder to extract the key features from speech signals. The HuBERT-equipped RVQs further decouple these features and hierarchically quantize them into discrete semantic and acoustic tokens. The discriminator enforces that the encoder and RVQs optimize their learned representations, aiming for comprehensive retention of the original audio's details. Through this structure, we can achieve effective decoupling of speech signals. The decoupled semantic and acoustic audio samples can be found on our demo page~\cite{safeear_demo}.  

\textbf{Encoder-Decoder Architecture.} To extract information-rich features $\mathbf{E}\in \mathbb{R}^{C\times T_n}$ from the raw audio $\mathbf{X}$, we follow the default configuration of Encodec~\cite{Encodec} to use the convolutional-based encoder-decoder architecture for detailed speech signal capture. As shown in Figure~\ref{fig:CDM}, although we remove the decoder during inference, it is vital for training to compel the audio codec to faithfully replicate the original audio, thus preserving the integrity and accuracy of the encoder's learned representation $\mathbf{E}$. In our design, we use the exponential linear unit (ELU) with layer normalization in each convolutional layer to enhance the nonlinear representations as well as the model's stability, and the decoder's structure mirrors that of the encoder. Moreover, to enhance the capability of semantic modeling, we replace Encodec's two-layer LSTM with a Bidirectional LSTM (Bi-LSTM). This modification allows for more precise capture of information across the audio feature space, producing as output a compound representation of essential semantic and acoustic properties of the raw audio for further processing. This design helps to improve the performance of RVQs feature decoupling.

\textbf{HuBERT-equipped RVQs for Decoupling.}
In CDM, we utilize Residual Vector Quantizers (RVQs) to effectively decouple semantic and acoustic tokens from the encoder's output $\mathbf{E}$. The RVQs employ cascaded vector quantization (VQ) layers, which project the input vector onto a predefined codebook to obtain a quantized representation. 
To effectively achieve decoupling, we have specifically designed and adjusted the RVQs, dividing it into two main parts: the semantic token part (VQ1) and the acoustic token part (VQ2$\sim$VQ8). 

In the semantic token part, we aim to modify the first quantizer (VQ1) to capture the semantic information from speech, serving a content-centric role. Specifically, we introduce a knowledge distillation approach, \textit{i.e.}, employing the well-established HuBERT~\cite{hsu2021hubert} as our semantic teacher of VQ1. Since HuBERT can well represent given speech as semantic-only features~\cite{mohamed2022self}, we employ the average representation across all HuBERT layers as the semantic supervision signal, which can encourage the semantic student VQ1 to learn a very close content representation via:
\begin{equation}\label{eq:distilliation}
    \mathcal{L}_{distill}=\frac{1}{T_n}\sum_{t=1}^{T_n}\log\sigma(\cos{(\mathbf{W}\cdot\mathbf{S}_t,\mathbf{H}_t)})
    \vspace{-5pt}
\end{equation}
where $\mathbf{S}_t$ is the VQ1 layer's quantized output and $\mathbf{H}_t$ is the semantic supervision signal at timestep $t$. $\cos(\cdot)$ is cosine similarity. $\sigma(\cdot)$ denotes sigmoid activation. $\mathbf{W}$ is the projection matrix.

Subsequently, in the acoustic token part, VQ1's semantic tokens $\mathbf{S}$ will be stripped away from the full-information encoder's output $\mathbf{E}$, resulting in purified acoustic information devoid of semantic information. These features are then passed to the subsequent seven quantizers (VQ2$\sim$VQ8), each further refining the acoustic information to enhance the feature representation of the sound. Through this layered and progressively refined processing, RVQs can handle complex sound data more efficiently. Ultimately, the outputs of all quantizers (VQ1$\sim$VQ8) are accumulated to form the input for the decoder. This accumulation process effectively recombines the semantic and acoustic information, enabling the decoder to reconstruct the original audio accurately. This design allows RVQs to effectively decouple audio content's semantic and acoustic properties while maintaining efficient encoding.
Please note that our design facilitates the cross-language decoupling, \textit{i.e.}, the VQ1 inherently takes the main information, so that despite our ``semantic teacher'' signal does not take the non-English corpus into account. \sys can also retain primary information in the VQ1 and the VQ2$\sim$VQ8 mainly describe speech details.

\textbf{Discriminator.}
Given the minimal differences between genuine and deepfake audio, our method is grounded in GAN-like adversarial training principles. By engaging discriminators and codec reconstruction in a mutually reinforcement iterative process, we force the encoder and RVQs to learn subtle speech representations, ensuring the preservation of fine-grained deepfake clues following feature decoupling.
Specifically, we adopt the same three discriminators as HiFi-Codec~\cite{yang2023hifi} that consist of the multi-scale STFT (MS-STFT)~\cite{chen2022fullsubnet+,chen2023inter}, the multi-periodic (MPD), and the multi-scale (MSD) discriminators. The MS-STFT discriminator analyzes complex-valued multi-scale STFTs, where real and imaginary parts are concatenated as input, to make spectrogram-level reconstruction results as similar as the original one~\cite{chen2022speech,chen2023gesper}. In contrast, the MPD and MSD focus on making the waveform-level reconstruction results as similar as the original one, \textit{i.e.}, the periodic elements and long-term patterns in the audio. These discriminators employ various sub-discriminators to analyze audio samples of different sizes and segments, ensuring the accuracy and integrity of the reconstructed audio. Due to the page limitations, we detail their objective functions as adversarial loss in Appendix~\ref{sec:CDM_loss}.

\color{black}
\subsection{Bottleneck \& Shuffle Layer}\label{sec:shuffle}
\begin{figure}[t]
    \centering
    \includegraphics[width=0.48\textwidth]{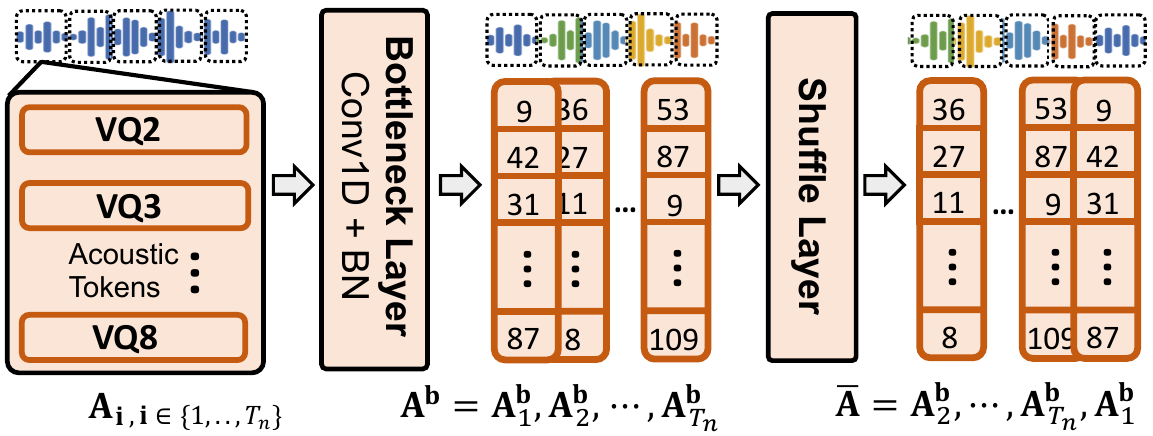}
    \caption{Bottlneck \& Shuffle layers (\ding{173}) of \sys.}
    \label{fig:shuffle}
\end{figure}

As shown in Figure~\ref{fig:shuffle}, the frontend CDM of \sys initially encodes waveform inputs into discrete acoustic tokens, $\mathbf{A}$, with each frame denoted as $\mathbf{A}_i$. The bottleneck layer aims to reduce the dimensions of acoustic tokens $\mathbf{A}$ from $\mathbb{R}^{7C\times T_n}$ to a more compact space $\mathbf{A}^{b} \in \mathbb{R}^{C\times T_n}$ by using 1D convolution and batch normalization. This layer serves a dual purpose: first, it enhances computational efficiency and reduces trainable parameters, facilitating subsequent layers to operate on a compact representation; second, it acts as a regularizer, avoiding over-fitting by limiting the amount of acoustic tokens and stabilizing it via batch normalization, before analyzed by the deepfake detector.

In addition to decoupling speech information, the shuffle layer serves to augment content protection by further scrambling the condensed acoustic tokens $\mathbf{A}^{b}$. As shown in Figure~\ref{fig:shuffle}, By randomly rearranging the elements across the temporal dimension $T_n$, this layer nullifies speech comprehension that is highly dependent on the temporal order of phonemes and words~\cite{Nature_contextual}. We empirically set a shuffling window of 1 second, corresponding to 50 frames, to obscure word-level intelligibility (as each token representation is extracted from a 20ms waveform). \blue{Thereby, the likelihood of attackers deciphering and correcting these sequences is extremely low, given the sheer number of possible permutations for a 4-second audio ($50!^4$, approximately $8.56 \times 10^{257}$, details are discussed in \S\ref{sec:disucssion})}. Our experiments also confirm the dual content protection by decoupling and shuffling, thwarting the advanced ASR techniques and human auditory analysis.

\subsection{Acoustic-only Deepfake Detector}\label{ssec:design_detection}
Recent studies \cite{ADD_survey,Rawformer} have indicated that the potential of Transformers in audio deepfake detection using full-information audio waveforms. In our scenario, however, the absence of semantic information combined with shuffling-induced acoustic patterns disorder (e.g., timbre and prosody) presents a unique challenge in detection.
\blue{To this regard, we develop a Transformer-based detector and determine its optimal 8 heads for Multi-Head Self-Attention (MHSA) mechanism~\cite{vaswani2017attention}. This configuration allows the model to more effectively engage in long-range feature interaction and dynamic spatial weighting.} It adeptly captures the slight differences between bonafide and deepfake audio. Moreover, it leverages parallel computation, allowing each attention head to independently process different aspects of the input feature space \cite{li2022efficient}. The aggregated features then form an attention spectrum, which is crucial for adaptively modulating features to more accurately detect deepfakes.

\begin{figure}[t]
    \centering
    \includegraphics[width=0.48\textwidth]{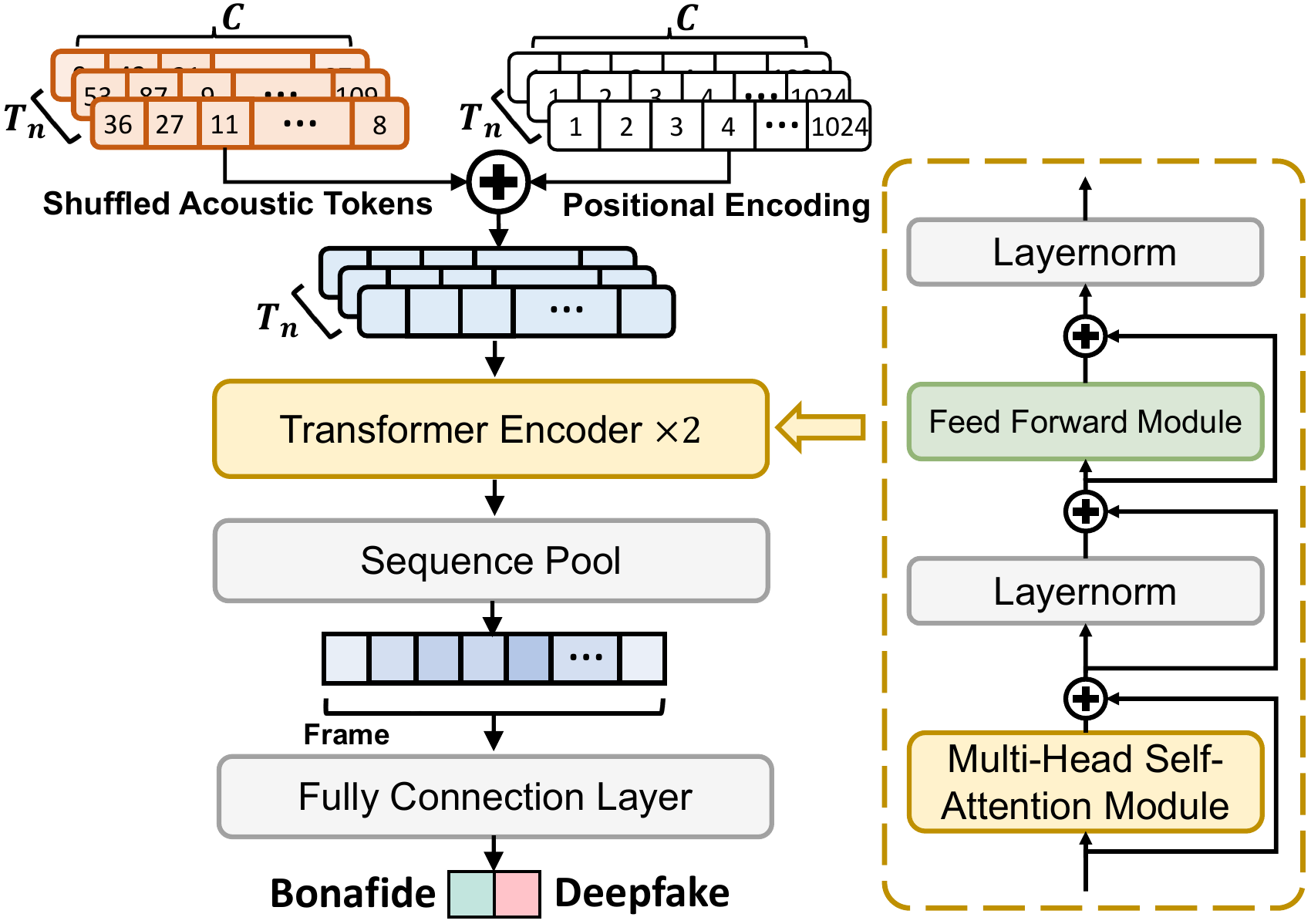}
    \caption{Acoustic-only deepfake detector (\ding{174}) of \sys.}
    \label{fig:bdm}
\end{figure}

As shown in Figure~\ref{fig:bdm}, we propose the Acoustic-only Deepfake Detector (ADD), which focuses on determining the genuineness of audio by analyzing only the shuffled acoustic tokens $\mathbf{\overline{A}}$.
Specifically, we first apply positional encoding to the sequence of shuffled acoustic tokens $\mathbf{\overline{A}}$ using sine and cosine alternating functions to enhance the MHSA modelling capabilities: 
\begin{align}
 \text{PE}(\mathbf{\overline{A}}, 2i)=\sin [\frac{\mathbf{\overline{A}}}{10000^{(\frac{2i}{C})}}];
 \text{PE}(\mathbf{\overline{A}}, 2i+1)=\cos [\frac{\mathbf{\overline{A}}}{10000^{(\frac{2i}{C})}}].
\end{align}
where $C$ denotes the token dimensions. We then feed $\mathbf{\overline{A}}$ into two sets of transformer encoders to process the sequence as a whole and capture global dependencies. Each set comprises two Feed-Forward Networks (FFNs), Multi-Head Self-Attention (MHSA), and Layernorm modules. The output from the Transformer encoders is finally directed to a fully connection layer, which determines whether the audio is a deepfake.

\subsection{Real-world Augmentation}\label{ssec:codec_augment}
It is noteworthy that the deepfake-and-bonafide gap in waveform can be degraded by real-world factors. Although studies have shown negligible differences in audible audio patterns across microphones \cite{li2023learning,hu2021speech}, we identify that codec transformations in real-world telecom channels pose a significant challenge in distinguishing genuine from deepfake audio. To address this challenge, we have strategically incorporated a few representative codecs into our training pipeline. These include OPUS~\cite{Opus}, known for its versatility and efficiency across audio types, and G.722~\cite{G.722}, renowned for high-quality voice transmission. We also utilize GSM for its widespread application in mobile communication, and both $\mu$-law and A-law~\cite{G.711} codecs, prevalent in North American, European, and international telephone networks. Additionally, we incorporate the MP3 codec~\cite{MP3}, a popular lossy compression technique in digital audio but introducing distortions and artifacts. Our diverse codecs integration strategy enables \sys to handle unique distortions each codec introduces and potentially generalize to more unseen coding technologies. The enhanced training process promote \sys maintains high accuracy and reliability in various real-world scenarios, where codec-induced variations are prevalent. Our augmentation excludes physical multi-channel information~\cite{zhang2021eararray,li2022use,li2023design} that is inapplicable to aid audio transmitted over the line.


\subsection{\sys Prototype}\label{ssec:prototype}
We have implemented a prototype of \sys using Pytorch 2.1~\cite{Pytorch}. During the training phase, we initially train \sys's codec-based decoupling model on LibriSpeech dataset \cite{librispeech} utilizing four RTX 3090 GPUs (NVIDIA), adhering to the procedure outlined in Equation~\ref{eq:all_loss}. We set the training epoch to 20. The maximum learning rate was set to $4\times 10^{-4}$, and the batch size of each GPU was 20. 
To better decouple the semantic and acoustic information of the input audio, we introduce multiple loss functions, including distillation loss $\mathcal{L}_{\text{distill}}$, reconstruction loss $\mathcal{L}_{\text{rec}}$, perceptual loss $\mathcal{L}_{\text{G}}$, and $\mathcal{L}_{\text{feat}}$ implemented via a discriminator, and RVQ commitment loss $\mathcal{L}_{\text{c}}$. 
The detailed loss functions are given in Appendix~\ref{sec:CDM_loss}. The CDM model's generator part is trained to optimize the following loss:
\begin{equation}\label{eq:all_loss}
    \mathcal{L}_\text{gen}= \lambda_{\text{d}} \mathcal{L}_{\text{distill}} + \lambda_{\text{r}} \mathcal{L}_{\text{rec}} + \lambda_{\text{G}} \mathcal{L}_{\text{G}} + \lambda_{\text{f}} \mathcal{L}_{\text{feat}} + \lambda_{\text{c}} \mathcal{L}_{\text{c}}
\end{equation}
where we set coefficients similar to HiFiGAN~\cite{kong2020hifi}, with specific values $\lambda_{\text{d}}=1, \lambda_{\text{r}}=1, \lambda_{\text{G}}=3, \lambda_{\text{f}}=3, \lambda_{\text{c}}=1$.

For the acoustic-only deepfake detector, we set the embedding dimensions to 1024, and the dropout rate in the model to 0.1. If not stated otherwise, we inverse \sys's acoustic token sequences within each 1s segment as the default shuffle approach. For the Transformer settings in the detector, we set the number of layers in the Transformer encoder to 2, the number of MHSA's heads to 8, and the positional encoding to be ``sinusoidal". We use BCE loss function and AdamW optimizer to optimize the detection model parameters with a learning rate of $3\times 10^{-4}$ and weight decay set to $1\times 10^{-4}$. Additionally, in each iteration of the training, we randomly extract a 4-second segment from speech samples and use one 3090 GPU. 

%% file: sections/exp_setup.tex
\section{Benchmark Construction} \label{sec:Benchmark}
We develop a comprehensive benchmark to evaluate different systems in terms of defending against \textit{deepfake adversaries} (DA), and \textit{content recovery adversaries} (CRA). The benchmark includes three deepfake datasets (\S\ref{ssec:deepfake_dataset}), two anti-content recovery datasets (\S\ref{ssec:content_dataset}).

\subsection{Comprehensive Deepfake Datasets}\label{ssec:deepfake_dataset}
To ensure our deepfake benchmark datasets cover a broad spectrum of TTS/VC techniques, we select the well-recognized ASVspoof 2019~\cite{ASVspoof2019_dataset} and ASVspoof 2021~\cite{ASVspoof2021_dataset} databases. Additionally, seeing the need for a cross-language deepfake benchmark~\cite{ADD_survey}, we establish a large-scale multilingual deepfake dataset using the CommonVoice corpus, in English, Chinese, German, French, and Italian~\cite{Commonvoice}. This dataset complements English-only ASVspoof 2019 and 2021 databases, forming a comprehensive benchmark (see Table~\ref{tab:dataset_statistic}). 

\subsubsection{ASVspoof 2019~\cite{ASVspoof2019_dataset}:} The ASVspoof 2019 LA subset comprises deepfake samples generated by 19 distinct TTS and VC systems. Adhering to the official guidelines, we use 6 deepfakes for training and the remaining 13 unseen deepfakes for testing. 

\subsubsection{ASVspoof 2021~\cite{ASVspoof2021_dataset}:} While sourced from ASVspoof 2019, the ASVspoof 2021 LA subset includes deepfake samples under more realistic conditions, where both bonafide and deepfake voice data are transmitted via telecom channels, \textit{e.g.}, VoIP. Its codec selection spans from traditional (\textit{e.g.}, a-law~\cite{G.711}) and modern IP streaming codecs (\textit{e.g.}, OPUS~\cite{Opus}) in use today, indicating mainstream usage. 

\subsubsection{Multilingual CVoiceFake:}
Current deepfake datasets are main\-ly single language-based and most of them are English deepfake audio datasets like ASVspoof 2019 \& 2021, and few of them encompass other languages, \textit{e.g.}, German or French. To facilitate cross-language deepfake detection research, we develop CVoiceFake, an extensive multilingual audio deepfake dataset comprising English, Chinese, German, French, and Italian, which is sourced from the widely used CommonVoice dataset~\cite{Commonvoice}. CVoiceFake also provides ground-truth transcriptions for each audio, making it an ideal benchmark for both deepfake detection (\S\ref{sec:eval1}) and content protection evaluation ({\S\ref{sec:eval2}). In alignment with deepfake techniques that adversaries likely use in real-world attacks, we employ five representative neural and digital signal processing (DSP) speech synthesis methods to yield deepfake samples, demo audio of which are available on website~\cite{safeear_demo}:
\begin{icompact}
\item \textbf{Parallel WaveGAN}~\cite{Parallel-WaveGAN}: As a non-autoregressive vocoder-based model, Parallel WaveGAN produces high-fidelity audio rapidly, ideal for efficient and quality deepfake generation.
\item \textbf{Multi-band MelGAN}~\cite{Multiband-MelGAN}: Multi-band MelGAN is a variant of MelGAN~\cite{MelGAN} that divides the frequency spectrum into sub-bands for faster and more stable multilingual vocoder training, enhancing the robustness and scalability of the dataset.
\item \textbf{Style MelGAN}~\cite{Style-MelGAN}: Style MelGAN is designed to capture fine prosodic and stylistic nuances of speech, making it particularly compelling for deepfake applications that require high levels of expressivity and variation in speech synthesis.
\item \textbf{Griffin-Lim}~\cite{Griffin-Lim}: This algorithm reconstructs waveforms from spectrograms using an iterative phase estimation method. Though less high-fidelity than neural vocoders, it serves as a traditional baseline for comparing deepfake generation.
\item \textbf{WORLD}~\cite{WORLD}: WORLD is a statistical parameter-based voice synthesis system that offers fine control over the spectral and prosodic features of the synthesized audio. Its fine manipulation is useful for crafting the nuanced variations needed in deepfake datasets.
\end{icompact}


In addition to utilizing high-fidelity vocoders for deepfake generation, we also implement MP3 compression on all genuine and synthesized speech samples. This step replicates the prevalent lossy media encoding used in social media platforms to enhance storage efficiency, thereby complementing the ASVspoof 2021's emphasis on the effects of transmission codecs. Overall, our benchmark integrates a comprehensive multilingual deepfake dataset, which features a range of deepfake generation methods and considers real-world encoding impacts.

\subsection{Anti-Content Recovery Datasets}\label{ssec:content_dataset}
Our benchmark also includes multilingual datasets to assess the performance of \sys in protecting user content privacy. 
The lack of ground-truth text references in ASVspoof challenge samples limits accurate evaluation of anti-\textit{content recovery adversaries} (CRA). We opt to utilize the widely adopted datasets in ASR tasks---LibriSpeech (English), and reuse CVoiceFake (English, Chinese, German, French, and Italian). Details are given in Table~\ref{tab:dataset_statistic}. 

\subsubsection{LibriSpeech \cite{librispeech}:} We utilize the train clean-100, clean-360, and other-500 subsets, totally extensive 960-hour corpus, for training CRA's ASR models. Then we test CRA's recovery ability using dev-clean, test-clean, and test-other subsets. These subsets offer a diverse range of accents and speaking styles in English, serving as a basis for evaluating the adversary's ability to reconstruct speech and compromise content privacy.

\subsubsection{Multilingual CVoiceFake:} We reuse our developed CVoiceFake dataset since it offers ground-truth transcriptions of each audio, and we employ their original uncompressed version. This presents an optimal condition for the CRA to infer speech content. \sys's successful privacy protection in this context highlights its robustness against CRA across diverse linguistic backgrounds.
\input{tables/dataset_single_col}

%% file: tables/dataset_single_col.tex
\begin{table}[]
\centering
\small
\renewcommand{\arraystretch}{1} 
\setlength{\tabcolsep}{4.5pt}
\setlength{\abovecaptionskip}{0pt}%
\setlength{\belowcaptionskip}{0pt}%
\caption{Statistics of benchmark datasets.}
\begin{tabular}{cclccc}
\toprule[1.2pt]
\textbf{Task$^\ddagger$} & \textbf{Dataset} & \multicolumn{1}{c}{\textbf{Char.$^\natural$}} & \textbf{Lang.$^\star$} & \textbf{Samples} & \textbf{Duration (s)} \\
\midrule[1pt]
T1 & ASVspoof 2019 & clean & En & 96,617 & 0.470$\sim$16.548 \\
\midrule
T1 & ASVspoof 2021 & telecom & En & 173,556 & 0.355$\sim$13.402 \\
\midrule
\multirow{5}{*}{\begin{tabular}[c]{@{}c@{}}T1\\ +\\ T2\end{tabular}} & \multirow{5}{*}{\begin{tabular}[c]{@{}c@{}}CVoiceFake\\ (Multilingual)\end{tabular}} & \multirow{5}{*}{media} & En & 257,581 & 0.972$\sim$10.692 \\
 &  &  & Cn & 254,116 & 1.512$\sim$19.656 \\
 &  &  & De & 239,127 & 1.476$\sim$11.124 \\
 &  &  & Fr & 284,351 & 0.792$\sim$11.808 \\
 &  &  & It & 219,718 & 0.792$\sim$14.112 \\
\midrule
T2 & Librispeech & clean & En & 289,503 & 1.285$\sim$34.955\\
\bottomrule[1.2pt]
\end{tabular}
\begin{tablenotes}[flushleft]
    \item[] \vspace{-2pt}\hspace{-2pt}\small 
    (1) $\ddagger$: T1 means Task 1, which serves as a benchmark to assess anti-deepfake adversary; T2 means Task 2, which serves as a benchmark to assess anti-content recovery adversary. (2) $\natural$: Char means the characteristics of the dataset, where ``telecom'' means using telecom codecs and ``media'' means using the MP3 codec for evaluating real-world factors. (3) $\star$: En: English, Cn: Chinese, De: German, Fr: French, and It: Italian.
\end{tablenotes}
\label{tab:dataset_statistic}
\end{table}

%% file: sections/evaluation.tex
\section{Evaluation: Deepfake Detection}\label{sec:eval1}
In this section, we focus on the \textbf{task 1 (T1)}: anti-\textit{deepfake adversary}, involving a comparative analysis of \sys against eight baselines across three deepfake benchmark datasets. We also investigate different impact factors, \textit{i.e.}, transmission codecs, deepfake techniques, and unseen-language deepfakes.

\subsection{Experiment Setup}
\textbf{Baselines.} We choose 8 representative baselines including end-to-end detectors---AASIST~\cite{AASIST}, RawNet 2~\cite{RawNet2}, and Rawformer~\cite{Rawformer}---take raw waveforms as input, as well as representative pipeline detectors---LFCC + SE-ResNet34~\cite{LFCC_SE-ResNet34}, LFCC + LCNN-LSTM~\cite{LCNN-LSTM}, LFCC + GMM~\cite{ASVspoof2021-baselines}, and CQCC + GMM~\cite{ASVspoof2021-baselines}. These baseline choice draws upon the recent state-of-the-art findings and official countermeasures provided by the ASVspoof challenge community. We also implement a frontend Wav2Vec2 feature-based system whose Transformer-based detector is configured the same as \sys for a fair comparison.

\noindent\textbf{Metrics.} We follow two standard metrics for audio deepfake detection~\cite{ASVspoof2019_report}. (1) \textit{Equal Error Rate }(EER): it characterizes the point at which the false acceptance rate equals the false rejection rate in deepfake detection; a system with lower EER exhibits more precise detection capability. (2) \textit{Tandem Detection Cost Function} (t-DCF): Unlike EER, it quantifies the cost-risk balance of false acceptances and false rejections, considering the prior probabilities of encountering bonafide versus deepfake utterances; a lower t-DCF indicates a better performance. Detailed formulations are in Appendix~\ref{appendix:t-DCF}.

\input{tables/1_ASVspoof_all}
\input{tables/2_multi_CVoiceFake}

\subsection{Overall Performance} 
We present the overall performance comparison of \sys with 8 baseline detectors, as detailed  in Table~\ref{tab:test_overall_asvspoof} for English ASVspoof 2019 and 2021, and in Table~\ref{tab:test_overall_CVoiceFake} for multilingual CVoiceFake. Note that for each baseline system, we have replicated and verified their performance, and herein report the official results.\\
\noindent\textbf{ASVspoof 2019 and 2021 (English).} Table~\ref{tab:test_overall_asvspoof} demonstrates that \sys outperforms the majority of baselines on these two datasets. In the ASVspoof 2019 dataset, \sys achieves a lower EER of 3.10\% than the average 4.90\% EER of all other baselines and a comparable t-DCF of 0.149. In the more challenging ASVspoof 2021 dataset, although we observe a general degradation, \sys's superiority is even more pronounced by achieving an EER of 7.22\% and t-DCF of 0.336, surpassing an average 11.07\% EER and 0.420 t-DCF across all baselines.
We make three key observations. Firstly, on ASVspoof 2019, four detection systems surpass the state-of-the-art 4.04\% EER reported in~\cite{ASVspoof2019_report}, \textit{i.e.}, AASIST, Rawformer, Wav2Vec2 + Transformer, and \sys. \blue{Notably, we supply acoustic-only tokens to other pipeline detectors, while the results demonstrate a marked degradation in performance: SE-ResNet34 decreases from 4.80\% to 6.09\%, LCNN-LSTM from 5.06\% to 10.41\%, and GMM from 8.09\% to 15.73\%. We envision that this decline is due to the classifier architectures being not designed for reliably extracting deepfake clues from shuffled and semantically-devoid tokens, indicating the effectiveness of \sys's tailored deepfake detector.}

On ASVspoof 2021, \sys outperforms most systems and exhibits comparable EER and t-DCF with Wav2Vec2 + Transformer, suggesting the effectiveness of \sys in resisting diverse audio deepfakes that are transmitted through varying channels.
Secondly, end-to-end models exhibit superior performance on ASVspoof 2019 due to their full leverage of speech information, enabling optimal speech representations for deepfake detection. However, they exhibit under-generalization on ASVspoof 2021, and raise privacy concerns due to their need of complete speech recordings.
Lastly, the Wav2Vec2-based system maintains consistent performance, likely due to its extensive pretraining on diverse audio inputs, offering a transferable speech representation. However, this advantage also presents a risk, because \textit{content recovery adversaries} could easily exploit such features for decoding intelligible content as we elaborate in Task 2 (\S\ref{sec:eval2}).\\
\noindent\textbf{CVoiceFake (Multiligual).}
Given the widespread misuse of deepfakes in the context of different languages, we compare \sys against above three top baseline systems: AASIST, Rawformer, and Wav2Vec2 + Transformer.
For a fair comparison, we randomly select 80\% speech samples from each language subset for training, reserving the remaining 20\% for testing. As shown in Table~\ref{tab:test_overall_CVoiceFake}, \sys achieves an average EER of 2.02\%, comparable to the performance of full-information-based AASIST and Rawformer, suggesting its multi-language detection ability. We consider Wav2Vec2's suboptimal performance on CVoiceFake is attributed to its incompatibility with excessively low MP3 bitrates like 48~kbit/sec~\cite{ASVspoof2021_dataset}, impeding its feature extraction, whereas \sys leverages robust neural codec architectures~\cite{Encodec} that maintain reliable acoustic tokens extraction even at low bitrates.

\subsection{Different Transmission Codecs}
Given the potential for fraudulent activities executing through diverse communication tools worldwide, we see the importance of robust detection across different telecom channels. For a fair comparison, we employ the identical real-world augmentation strategy as detailed in \S\ref{ssec:codec_augment} to train each detector, as shown in Table~\ref{tab:test_asvspoof_codec}. Then we evaluate the impact of telecom channels using 6 representative codecs officially set in the ASVspoof 2021 challenge, including a-law, G722, GSM, OPUS, unknown, $\mu$-law, and a no codec scenario for baseline comparison. We observe despite there are slight performance gap against Rawformer, \sys is on par with Wav2Vec2 across most codecs and generally outperforms the end-to-end AASIST. Another finding is a consistent decline in performance when detecting unknown codecs. This decline is likely due to the sequential compressions these codecs undergo across multiple telecom channels, resulting in a more significant loss of signal fidelity compared to mainstream codecs. 
\input{tables/3_transmission_codec}

\subsection{Different Deepfake Techniques}
We compare \sys with baselines on a spectrum of prevalent deepfake vocoders and analyzes the individual performance in Table~\ref{tab:test_CVoiceFake_vocoder}.
\sys shows remarkable vocoder-agnostic detection capability across all tested cases, hitting overall 2.02\% comparable to AASIST and Rawformer and surpassing Wav2Vec2 significantly. In real-life scenarios, \textit{deepfake adversaries} are likely to employ advanced neural vocoders, such as Multiband-MelGAN, Parallel-WaveGAN, and Style-MelGAN to produce highly convincing synthetic speech. \sys can even hit 0.61\% EER, highlighting its efficacy to thwart sophisticated deepfake methods. We validate higher EERs in the classical deepfake technique, Griffin-Lim, is caused by that the attention of model is trained to focus on minor artifacts existed in other four advanced vocoders, thus leading to minor degradation. For instance, our further individual training on Griffin-Lim, denoting \sys can detect it with 2.01\% EER. We envision that a holistic system can ensemble different detectors trained on individual deepfake technologies.
\input{tables/4_vocoder}

\subsection{Unseen-Language Deepfake Detection} 
With a numerous user base engaging in virtual communications daily, \sys may encounter deepfake speech spoken in unseen languages. We consider a challenging scenario where \sys's transformer detector is trained only in one language and then identifies deepfake audios across all five languages. Table~\ref{tab:test_cross_language} demonstrates that without a comprehensive training with multi-language data, the performance of the Transformer-based detector degrades. For instance, the detector trained on English obtains 15.92\% EER on French and 9.70\% average EER across five languages, while the optimal average EER is down to 2.02\% as shown in Table~\ref{tab:test_overall_CVoiceFake}. We also find that the choice of training language impacts to a certain degree. For instance, the detector trained on Chinese data achieves an average EER of 5.19\%, lower than other settings, like 9.70\% (English). These findings highlight the necessity for more multilingual datasets to develop practical deepfake detection approaches.

\input{tables/9_detect_cross_language}

\section{Evaluation: Content Protection}\label{sec:eval2}
In this section, we focus on the \textbf{task 2 (T2)}: anti-\textit{content recovery adversaries}. We consider three kinds of content recovery adversaries, \textit{i.e.}, \textit{naive} (CRA1), \textit{knowledgeable} (CRA2), and \textit{adaptive} (CRA3), with different knowledge and capabilities.
\subsection{Experiment Setup}
\hspace{0.4cm}\textbf{Adversary Definition.} We define three content recovery adversaries that pose threats to \sys:
\begin{icompact}
    \item \textit{Naive content recovery adversary} (CRA1): The adversary lacks knowledge of \sys's internal parameters. However, CRA1 can emulate user interactions with \sys to input known speech, thereby acquiring a substantial dataset of pairs of \sys's tokens and ground-truth text. In our evaluation, CRA1 can acquire an extensive 960-hour Librispeech corpus to train advanced ASR models for recovering text from received tokens.
    \item \textit{Knowledgeable content adversary} (CRA2): In contrast, CRA2 is assumed to have the knowledge of \sys's algorithm and can replicate its decoder. With this knowledge, CRA2 does not need to collect numerous data for ASR training. Instead, CRA2 can reconstruct speech waveform from an individual speech sample's acoustic tokens and apply advanced ASR models or human auditory analysis for recognizing content.
    \item \textit{Adaptive content adversary} (CRA3): We assume this most advanced adversary can even deduce the shuffled order of a given token sequence and rectify it with a few attempts, allowing CRA3 to derive the original acoustic token sequence and then recover content as CRA2 does.
\end{icompact}

\textbf{Baselines.} We envision that content recovery adversaries can employ 7 state-of-the-art ASR systems, including local and commercial ASRs. For CRA1, we compare the content recovery efficacy based on \sys and other inputs, leveraging the leading Bi-LSTM~\cite{graves2013hybrid} and Conformer~\cite{gulati2020conformer} ASR architectures. For CRA2, we utilize the well-recognized local Wav2Vec2~\cite{speechbrain} and 4 commercial ASRs~\cite{tecentasr,xfasr,azureasr,amazonasr} to compare \sys and other from CRA2's reconstructed speech waveforms as inputs. For CRA3, we keep the same setting as CRA2 yet this most advanced adversary can rectify shuffled acoustic tokens before speech reconstruction.


\textbf{Metrics.} \textit{(1) Word/Character Error Rate} (WER/CER): they measure the accuracy of content recovery from processed audio by indicating the proportion of words or characters incorrectly transcribed by an ASR system. A higher WER/CER denotes a better privacy-preserving ability against content recovery attacks. \blue{Note that WER can exceed 100\% because its upper bound is $max(N1,N2)/N1$~\cite{morris2004and}, where N1 and N2 are the number of words in ground-truth and ASR transcription.}
\textit{(2) Short-Time Objective Intelligibility} (STOI)~\cite{taal2011algorithm}: it indicates speech signal intelligibility with its range quantified from 0 to 1 to represent the percentage of words that are correctly understood. A lower STOI means a better privacy-preserving ability. \textit{(3) Subjective Assessment}: we conduct a user study in \S\ref{ssec:user_study} that includes three sub-metrics---ASR effectiveness, human intelligibility, and human WER. 

\subsection{\textit{Anti-Naive Adversary} (CRA1)}\label{ssec:CRA1}
In this part, we assess \sys's efficacy in multi-language content protection against recovery attacks (CRA1). These adversaries can gather shuffled acoustic tokens and corresponding ground-truth text pairs from \sys to train advanced Bi-LSTM and Conformer models. Given that advanced end-to-end detectors like AASIST and Rawformer, which take raw waveforms as inputs, alongside the Wav2Vec2-based pipeline detector, we include both input types for evaluation. Additionally, \sys's capacity for semantic-acoustic decoupling is evaluated, using its semantic tokens as a baseline for comparison.	

\begin{figure}[t]
    \centering
    \includegraphics[width=0.4\textwidth]{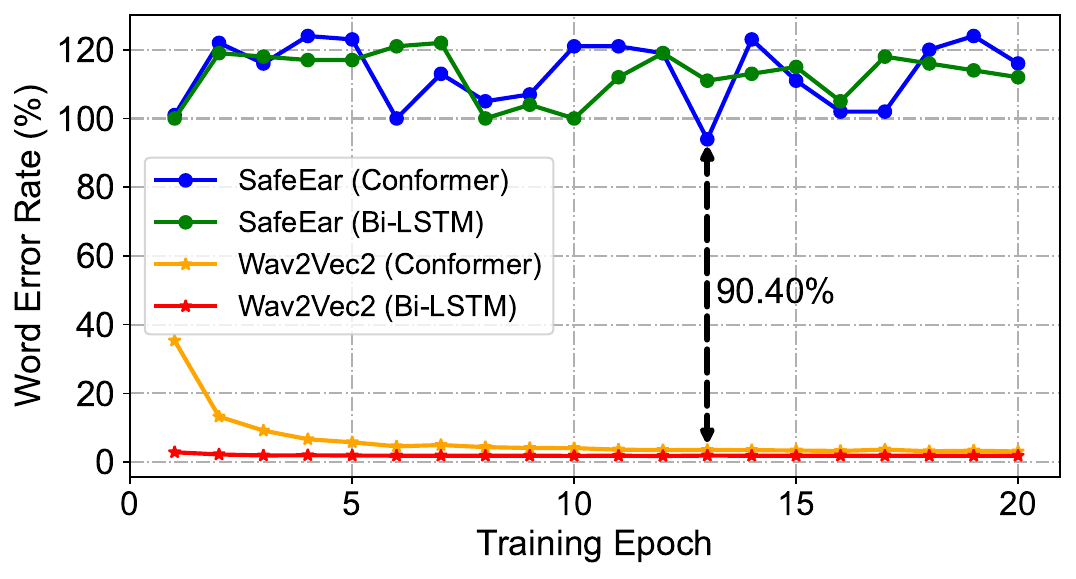}
    \caption{WER curves validated on the dev-clean set during training (CRA1).}
    \label{fig:eval_wer_dev_clean}
\end{figure}

\textbf{CRA1---English Content Protection.}
Table~\ref{tab:test_libri_token_ASR} demonstrates that CRA1 can easily infer users' speech content when receiving raw waveform and Wav2Vec2 feature inputs, with all WERs below 10.46\%. Bi-LSTM and Conformer separately transcribe Wav2Vec2 and waveforms better, with minimal 1.78\% and 2.55\% WERs. 
As for semantic tokens, all WERs below 19.61\% and a minimum WER of 6.68\% indicates that \sys well decouples semantic information from speech. In contrast, the acoustic tokens effective in deepfake detection, yet inapplicable for conversion back into intelligible content, even when CRA1 trains both ASR models using 960-hour Librispeech dataset over multiple epochs. As shown in Figure~\ref{fig:eval_wer_dev_clean}, during the training of ASR models based on acoustic tokens, the validation WER curves of \sys remain high and do not converge, keeping 90.40\% WER higher than the Wav2Vec2-based system, highlighting \sys's resilience against content recovery attacks. Finally, the WERs and CERs are still too high: 93.93$\sim$106.2\% and 72.74$\sim$97.12\%, respectively, far surpassing the unacceptable WER threshold of over 45\% as reported in~\cite{WER_report}. The results of our user study (see \S\ref{ssec:user_study}) also confirms that these ASR-transcribed text are unintelligible.

\input{tables/5_librispeech_ASR}


\input{tables/6_CommonVoice}
\textbf{CRA1---Unseen Language Content Protection.}
As \sys's semantic-acoustic decoupling ability derives from the English-based HuBERT teacher, we evaluate its effectiveness in protecting unseen-language content, including Chinese, German, French, and Italian. 
We keep Wav2Vec2 with the lowest WER in Table~\ref{tab:test_libri_token_ASR} as a baseline comparison. Table~\ref{tab:test_commonvoice_token_ASR} shows that CRA1 can train Wav2Vec2-based ASRs~\cite{speechbrain} to obtain acceptable WERs with audio recorded in non-ideal conditions, while \sys well impedes adversaries in training usable ASRs. This is evidenced by all WERs exceeding 94.82\%, suggesting a substantial error rate in recovered information. We attribute the zero-shot speech disentanglement ability to two reasons: First, neural codec models possess the language-agnostic properties for compression and decompression, making them suitable for various instant communication platforms. \sys, built on this foundation, succeeds cross-language ability. Second, as detailed in \S\ref{sec:CDM}, the RVQs architecture of \sys's frontend CDM facilitates primary information retained in its VQ1, and the VQ2$\sim$VQ8 mainly describe speech details like prosody and timbre. Third, we consider that the shuffle operation also interferes ASRs to transcribe.

\input{tables/7_1_Librispeech_API}
\subsection{\textit{Anti-Knowledgeable Adversary} (CRA2)}\label{ssec:CRA2}
In this part, we evaluate the resistance of \sys against \textit{knowledgeable content adversaries} (CRA2), who can reconstruct received tokens into speech waveforms and employ off-the-shelf ASR models or even human auditory to analyze speech content across different languages.


\textbf{CRA2---English Content Protection.}
To comprehensively evaluate CRA2's ability to recover content, we select the best local ASR, \textit{i.e.}, Wav2Vec2~\cite{facebook_wav2vec2} and four commercial ASR APIs out of multiple off-the-shelf candidates.
As illustrated in Table~\ref{tab:test_speech_libri_ASR}, the original speech waveforms serve as an optimal baseline, based on which, CRA2 can obtain a low transcription WERs of 3.15\% and 7.68\% on two subsets. 
In the ``Coded'' reference group where audio samples are processed by the representative telecom codec---OPUS, CRA2 maintains comparable WERs as low as 3.82\% and 11.83\%, respectively. This results confirms that CRA2 can easily eavesdrop speech content within virtual calls or meetings despite distortion exists.  
In contrast, \sys significantly safeguards the actual speech content by shuffled acoustic tokens, resulting in an average WER above 99.94\%, a level too high for adversaries to meaningfully interpret the content. Additionally, as shown in Table~\ref{tab:STOI}, the STOI metric, used for assessing the objective intelligibility of CRA2's reconstructed speech samples, further substantiate inefficacy of CRA2 in understanding data anonymized by \sys, with values of 0.0018 and 0.0015, significantly lower than 0.8698 and 0.8719 of ``Coded''.

\input{tables/8_commonvoice_API}
\textbf{CRA2---Unseen Language Content Protection.}
CRA2 may employ established ASR models for different languages to conduct content recovery across diverse linguistic contexts. We report \sys's effectiveness in protecting content in unseen languages against CRA2 in Table~\ref{tab:test_speech_commonvoice_ASR}, omitting the coded setting due to its results being very close to the original audio. Results indicate that CRA2 can recover meaningful content from multilingual original audio with slightly higher WER due to audio's lower quality. However, \sys still safeguards content privacy, maintaining all WERs above 90.89\% and averaging 102.63\% across five ASR models. As shown in Table~\ref{tab:STOI}, the objective STOI values for \sys all approach 0, ranging between 0.0031 and 0.0106. In contrast, the STOI values for the ``Coded'' condition consistently exceed 0.7326. This remarkable contrast confirms the efficacy of \sys in unseen-language content protection. Moreover, these results conform with the subjective intelligibility of our user study (see \S\ref{ssec:user_study}).
\input{tables/12_reconstructed_STOI}

\subsection{\textit{Anti-Adaptive Adversary} (CRA3)}\label{ssec:CRA3}
\begin{figure}[t]
\centering
    \subfloat[WER/CER comparison on the Librispeech dataset]{
        \includegraphics[width = 0.4\textwidth]{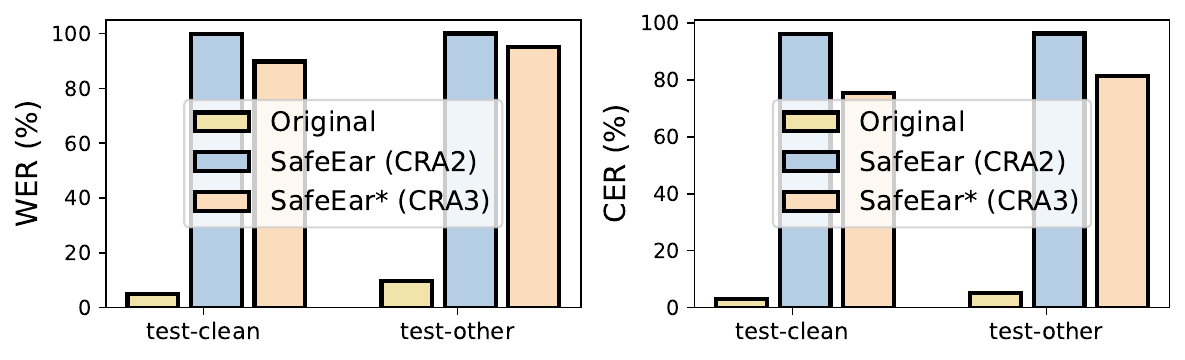}
    }
    \hfill
    \subfloat[WER comparison on the CVoiceFake dataset]{
        \includegraphics[width = 0.4\textwidth]{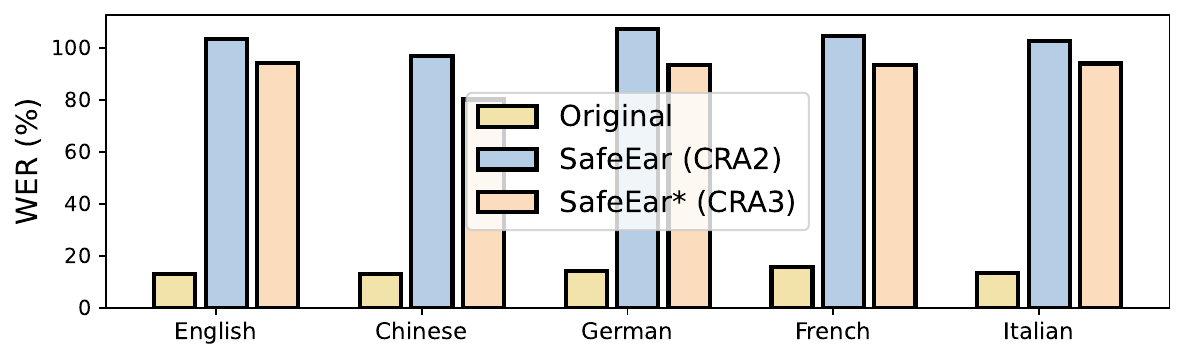}
    }
\caption{Adaptive adversary's (CRA3) recovery performance on different datasets compared with CRA2.}
\label{fig:adaptive_ASR_bar}
\end{figure}

In this part, we explore whether \sys can safeguard speech content from recovery by the most adaptive adversary (CRA3). This evaluation also serves as an ablation study that examines the standalone content protection ability of acoustic tokens. CRA3 adversaries are distinguished from CRA1 and CRA2 by their ability to rectify the correct temporal sequence of acoustic tokens $\mathbf{A}$, denoted as ``SafeEar*'', even after random shuffling to $\mathbf{\overline{A}}$. For direct comparison, we put above three types of audio samples on our website~\cite{safeear_demo}. As shown in Figure~\ref{fig:adaptive_ASR_bar}, an overall decrease in WER/CERs compared to \sys (CRA2) is observed, indicating CRA3's slight improvement in content comprehension. However, these rates remain too high to comprehend, due to acoustic tokens' devoid of semantic information. 
Furthermore, we envision that an adaptive adversary would repeatedly listen to the correct-order speech to interpret it. To explore this, we have established a user study in \S\ref{ssec:user_study}, including three aspects of subjective assessment.

\subsection{User Study}\label{ssec:user_study}
To validate \sys's content protection against machine-based and human auditory analysis, we conduct a user study, which is approved by the Institutional Review Board (IRB) of our institute. 

\textbf{Setup.} We have recruited 68 participants, aged 21$\sim$35 years and comprising 51 males and 17 females with bilingual proficiency in English and Chinese. Our user study includes two sets of questions:
(1) \textit{ASR effectiveness}. To evaluate whether human adversaries can extract meaningful information from content transcribed by both self-trained and off-the-shelf ASR models, we set a metric, named ASR effectiveness. Participants are asked to rate on a scale of 1$\sim$10 points (1 indicating no correlation, and 10 indicating exact match) their ability to deduce the original text from machine-transcribed results. (2) \textit{Intelligibility \& Human WER:} To assess whether \sys can shield speech reconstruction from human auditory analysis. Participants are asked to listen to audio samples and rate their clarity on a scale of 1 to 10 (1 being entirely unintelligible, and 10 being crystal clear). Subsequently, they manually transcribed the speech content for human-ear WER calculation. Participants were required to act themselves as content recovery adversaries (CRA), and answered all questions under a quiet environment to better emulate the optimal content recovery performance.

\begin{figure}[t]
    \centering
    \includegraphics[width=0.48\textwidth]{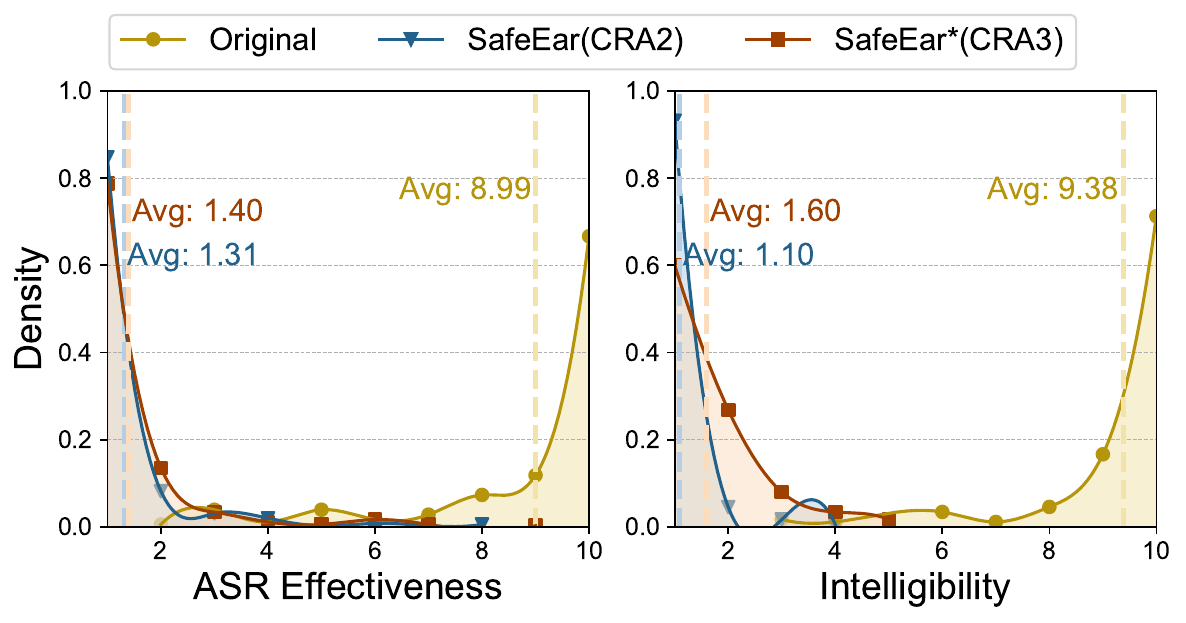}
    {\small{
        \begin{tabular}{cccc}
        \toprule[1.2pt]
        \textbf{Human WER$\uparrow$} & Original & SafeEar (CRA2) & SafeEar* (CRA3) \\ \midrule
        English & 24.45\% & 98.31\% & 96.37\% \\
        Chinese & 11.32\% & 99.75\% & 98.92\% \\
        \bottomrule[1.2pt]
        \end{tabular}
        }
    }
    \caption{Results of the user study: ASR effectiveness, Intelligibility, and Human WER metrics vary with three types of speech---Original, \sys (CRA2), and SafeEar* (CRA3).}
    \label{fig:user_study}
\end{figure}

\textbf{Results.} Figure~\ref{fig:user_study} illustrates the findings on the three pivotal metrics. We categorized and analyzed the results based on different levels of test speech sample reconstruction: Original, \sys (CRA2), and SafeEar* (CRA3). 
In line with above experiments, original speech samples represented baseline performance of existing deepfake detectors without content privacy protection. The study reveals that participants can discern actual content from ASR-transcribed text, evidenced by high average scores of 8.99 in ASR effectiveness and 9.38 in intelligibility. Manual transcription attempts yield acceptable 24.45\% and 11.32\% WER in English and Chinese, respectively, where the accuracy is slightly affected by the variance of individual auditory abilities.
In contrast, metrics significantly drops under \sys protection in CRA2 and CRA3 scenarios. As speech samples are reconstructed from shuffled acoustic-only information in CRA2 cases, participants struggled to deduce content from meaningless transcriptions, resulting in average scores of 1.31 in ASR effectiveness and 1.10 in intelligibility, with human WERs soaring to 98.31\% and 99.75\%.
Although adversaries may reconstruct the acoustic tokens with correct order into speech (CRA3), participant responses confirm the failure of both machine and human auditory analysis, with negligible improvements (1.40 in ASR effectiveness, 1.60 in intelligibility, and persistently high WERs).
Consequently, \sys well safeguards content privacy against both machine and human auditory analysis.

%% file: tables/1_ASVspoof_all.tex
\begin{table}[t!]\centering
\small
\renewcommand{\arraystretch}{1} 
\setlength{\tabcolsep}{2pt}
\setlength{\abovecaptionskip}{0pt}%
\setlength{\belowcaptionskip}{0pt}%
\caption{[T1] Overall Performance of \sys compared with baselines on ASVspoof 2019 \& 2021 datasets.}
\begin{tabular}{llcccc}
\toprule[1.2pt]
\multicolumn{1}{c}{\multirow{2}{*}{\textbf{Type$^\ddagger$}}} & \multicolumn{1}{c}{\multirow{2}{*}{\textbf{Method}}} & \multicolumn{2}{c}{\textbf{ASVspoof 2019}} & \multicolumn{2}{c}{\textbf{ASVspoof 2021}} \\
\multicolumn{1}{c}{} & \multicolumn{1}{c}{} & \textbf{EER (\%)}$\downarrow$ & \textbf{t-DCF}$\downarrow$ & \textbf{EER (\%)}$\downarrow$ & \textbf{t-DCF}$\downarrow$ \\
\midrule[1pt]
\multirow{3}{*}{E2E} & AASIST & 1.20 & 0.034 & 9.15 & 0.437 \\
 & RawNet 2 & 5.64 & 0.130 & 9.50 & 0.426 \\
 & Rawformer & 1.05 & 0.034 & 8.72 & 0.397 \\
\midrule
\multirow{6}{*}{pipe} & LFCC + SE-ResNet34 & 4.80 & 0.098 & 10.39 & 0.355 \\
 & LFCC + LCNN-LSTM & 5.06 & 0.156 & 9.26 & 0.345 \\
 & LFCC + GMM & 8.09 & 0.212 & 19.30 & 0.576 \\
 & CQCC + GMM & 9.57 & 0.237 & 15.62 & 0.497 \\
 & Wav2Vec2 + Transformer & 3.82 & 0.184 & 6.64 & 0.330 \\
 & \cellcolor{lightgray}\textbf{SafeEar (Ours)} & \cellcolor{lightgray}3.10 & \cellcolor{lightgray}0.149 & \cellcolor{lightgray}7.22 & \cellcolor{lightgray}0.336\\ 
 
\bottomrule[1.2pt]
\end{tabular}
\begin{tablenotes}[flushleft]
    \item[] \vspace{-2pt}\hspace{-2pt}\small 
    $\ddagger$: E2E: An end-to-end detector takes speech's raw waveform as input; pipe: A pipeline detector employs a frontend module to extract speech representation, such as LFCC, CQCC, and Wav2Vec2, then feeding it to a backend classifier like SE-ResNet34, LCNN-LSTM, GMM, and Transformer.
\end{tablenotes}
\label{tab:test_overall_asvspoof}
\end{table}

%% file: tables/2_multi_CVoiceFake.tex
\begin{table}[t!]\centering
\small
\renewcommand{\arraystretch}{1} 
\setlength{\tabcolsep}{3pt}
\setlength{\abovecaptionskip}{0pt}%
\setlength{\belowcaptionskip}{0pt}%
\caption{[T1] Overall Performance of \sys compared with baselines on the CVoiceFake dataset.}
\begin{tabular}{@{}lcccccc@{}}
\toprule[1.2pt]
\multicolumn{1}{c}{\multirow{2}{*}{\textbf{Method}}} & \multicolumn{6}{c}{\textbf{CVoiceFake EER (\%) $\downarrow$}} \\
\multicolumn{1}{c}{} & \multicolumn{1}{c}{English} & \multicolumn{1}{c}{Chinese} & \multicolumn{1}{c}{German} & \multicolumn{1}{c}{French} & \multicolumn{1}{c}{Italian} & \multicolumn{1}{c}{Average} \\ 
\midrule[1pt]
AASIST & 1.63 & 1.50 & 1.63 & 2.79 & 1.89 & 1.89  \\
Rawformer & 1.13 & 1.50 & 1.13 & 1.85 & 0.81 & 1.28  \\
Wav2Vec2 & 12.33 & 10.17 & 12.33 & 13.59 & 9.45 & 11.57  \\
\rowcolor{lightgray}
\textbf{SafeEar (Ours)} & 2.01 & 1.63 & 1.77 & 2.80 & 1.89 & 2.02  \\ 
\bottomrule[1.2pt]
\end{tabular}
\begin{tablenotes}[flushleft]
    \item[] \vspace{-2pt}\hspace{-2pt}\small 
    $\ddagger$: Wav2Vec2: simplified for Wav2Vec2 + Transformer.
\end{tablenotes}
\label{tab:test_overall_CVoiceFake}
\end{table}

%% file: tables/3_transmission_codec.tex
\begin{table}[t!]\centering
\small
\renewcommand{\arraystretch}{1} 
\setlength{\tabcolsep}{3pt}
\setlength{\abovecaptionskip}{0pt}%
\setlength{\belowcaptionskip}{0pt}%
\caption{[T1] Comparison of \sys and baselines in detecting deepfakes transmitted via different channels.}
\begin{tabular}{@{}lccccccc@{}}
\toprule[1.2pt]
\multicolumn{1}{c}{\multirow{2}{*}{\textbf{Method}}} & \multicolumn{7}{c}{\textbf{ASVspoof 2021 EER (\%) $\downarrow$}} \\
\multicolumn{1}{c}{} & a-law & G.722 & GSM & OPUS & unknown & $\mu$-law & / \\
\midrule[1pt]
AASIST & 7.17 & 10.07 & 8.15 & 19.86 & 17.18 & 7.17 & 8.31 \\
Rawformer & 2.64 & 2.28 & 3.91 & 3.23 & 5.73 & 2.5 & 2.36 \\
Wav2Vec2 & 4.89 & 4.39 & 6.16 & 4.28 & 6.5 & 4.46 & 4.04 \\
\rowcolor{lightgray}
\textbf{SafeEar (Ours)} & 6.13 & 4.35 & 8.19 & 4.96 & 9.74 & 6.25 & 4.06\\ 
\bottomrule[1.2pt]
\end{tabular}
\label{tab:test_asvspoof_codec}
\end{table}

%% file: tables/4_vocoder.tex
\begin{table}[t!]\centering
\small
\renewcommand{\arraystretch}{1} 
\setlength{\tabcolsep}{1.5pt}
\setlength{\abovecaptionskip}{0pt}%
\setlength{\belowcaptionskip}{0pt}%
\caption{[T1] Comparison of \sys and baselines in detecting deepfakes created by different synthetic techniques.}
\begin{tabular}{@{}lcccccc@{}}
\toprule[1.2pt]
\multicolumn{1}{c}{\multirow{3}{*}{\textbf{Technique}}} & \multicolumn{6}{c}{\textbf{CVoiceFake EER (\%) $\downarrow$}} \\
\multicolumn{1}{c}{} & Overall & \begin{tabular}[c]{@{}c@{}}Griffin\\ Lim\end{tabular} & WORLD & \begin{tabular}[c]{@{}c@{}}Multiband\\ MelGAN\end{tabular} & \begin{tabular}[c]{@{}c@{}}Parallel\\ WaveGAN\end{tabular} & \begin{tabular}[c]{@{}c@{}}Style\\ MelGAN\end{tabular} \\
\midrule[1pt]
AASIST & 1.89 & 2.88 & 1.03 & 0.99 & 0.70 & 1.46 \\
Rawformer & 1.28 & 2.27 & 1.29 & 0.52 & 0.57 & 0.96 \\
Wav2Vec2 & 11.57 & 23.64 & 7.78 & 7.04 & 8.98 & 6.24 \\
\rowcolor{lightgray}
\textbf{SafeEar (Ours)} & 2.02 & 3.68 & 0.99 & 0.76 & 0.61 & 1.37 \\
\bottomrule[1.2pt]
\end{tabular}
\label{tab:test_CVoiceFake_vocoder}
\end{table}

%% file: tables/9_detect_cross_language.tex
\begin{table}[t!]\centering
\small
\renewcommand{\arraystretch}{1} 
\setlength{\tabcolsep}{3pt}
\setlength{\abovecaptionskip}{0pt}%
\setlength{\belowcaptionskip}{0pt}%
\caption{[T1] Unseen language Detection Analysis.}
\begin{tabular}{lcccccc}
\toprule[1.2pt]
\multirow{2}{*}{\textbf{\sys}} & \multicolumn{6}{c}{\textbf{CVoiceFake EER (\%) $\downarrow$}} \\
 & English & Chinese & German & French & Italian & Average \\
\midrule[1pt]
English & \textbf{5.05} & 10.36 & 3.94 & 15.92 & 13.25 & 9.70 \\
Chinese & 6.68 & \textbf{2.75} & 5.42 & 6.45 & 4.65 & \textbf{5.19} \\
German & 5.98 & 9.07 & \textbf{1.33} & 14.76 & 11.93 & 8.61 \\
French & 11.62 & 6.56 & 9.87 & \textbf{6.56} & 6.89 & 8.30 \\
Italian & 7.81 & 4.54 & 7.40 & 6.06 & \textbf{3.57} & 5.88 \\
\bottomrule[1.2pt]
\end{tabular}
\label{tab:test_cross_language}
\end{table}

%% file: tables/5_librispeech_ASR.tex
\begin{table}[t!]\centering
\small
\renewcommand{\arraystretch}{1} 
\setlength{\tabcolsep}{2pt}
\setlength{\abovecaptionskip}{0pt}%
\setlength{\belowcaptionskip}{0pt}%
\caption{[T2] English (Seen language) content protection against \textit{naive adversary}'s recovery attacks (CRA1).}
\begin{tabular}{llcccc}
\toprule[1.2pt]
\multirow{2}{*}{\begin{tabular}[c]{@{}c@{}}\textbf{ASR}\\ \textbf{Architecture}\end{tabular}} & \multicolumn{1}{c}{\multirow{2}{*}{\textbf{Input$\natural$}}} & \multicolumn{2}{c}{\textbf{Libri. dev-clean}} & \multicolumn{2}{c}{\textbf{Libri. test-clean}} \\
 &  & WER (\%)$\uparrow$ & CER (\%)$\uparrow$ & WER (\%)$\uparrow$ & CER (\%)$\uparrow$ \\
\midrule[1pt]
\multirow{4}{*}{Bi-LSTM} & Waveform & 10.01 & 3.15 & 10.46 & 3.40 \\
 & Wav2Vec2 & 1.78 & 0.48 & 1.99 & 0.52 \\ 
 & Semantic & 19.03 & 5.79 & 19.61 & 5.84 \\
 & \cellcolor{lightgray}\textbf{\sys} & \cellcolor{lightgray}\textbf{100.2} & \cellcolor{lightgray}\textbf{94.85} & \cellcolor{lightgray}\textbf{101.4} & \cellcolor{lightgray}\textbf{97.12} \\
\midrule
\multirow{4}{*}{Conformer} & Waveform & 4.69 & 1.79 & 2.55 & 0.86 \\
 & Wav2Vec2 & 3.09 & 1.05 & 2.25 & 0.82 \\
 & Semantic & 11.64 & 4.92 & 6.68 & 3.11 \\
 & \cellcolor{lightgray}\textbf{\sys} & \cellcolor{lightgray}\textbf{93.93} & \cellcolor{lightgray}\textbf{72.74} & \cellcolor{lightgray}\textbf{106.2} & \cellcolor{lightgray}\textbf{78.76}\\
\bottomrule[1.2pt]
\end{tabular}
\begin{tablenotes}[flushleft]
    \item[] \vspace{-2pt}\hspace{-2pt}\small 
    $\natural$: Semantic means $\mathbf{S}$ from VQ1; \textbf{\sys} means acoustic tokens (VQ2$\sim$VQ8) goes through bottleneck \& shuffle layer as $\mathbf{\overline{A}}$.
\end{tablenotes}
\label{tab:test_libri_token_ASR}
\end{table}

%% file: tables/6_CommonVoice.tex
\begin{table}[t!]\centering
\small
\renewcommand{\arraystretch}{1} 
\setlength{\tabcolsep}{1pt}
\setlength{\abovecaptionskip}{0pt}%
\setlength{\belowcaptionskip}{0pt}%
\caption{[T2] Multilingual (Unseen language) content protection against \textit{naive adversary}'s recovery attacks (CRA1).}
\begin{tabular}{llccccc}
\toprule[1.2pt]
\multicolumn{1}{c}{\multirow{2}{*}{\textbf{\begin{tabular}[c]{@{}c@{}}ASR\\ Architecture\end{tabular}}}} & \multicolumn{1}{c}{\multirow{2}{*}{\textbf{Input}}} & \multicolumn{5}{c}{\textbf{CVoiceFake WER (\%) $\uparrow$}} \\
\multicolumn{1}{c}{} & \multicolumn{1}{c}{} & English & Chinese & German & French & Italian \\
\midrule[1pt]
\multirow{2}{*}{Conformer} & Wav2Vec2 & 15.69 & 19.03 & 8.93 & 10.24 & 8.38 \\
 & \cellcolor{lightgray}\textbf{\sys} & \cellcolor{lightgray}98.23 & \cellcolor{lightgray}94.82 & \cellcolor{lightgray}108.2 & \cellcolor{lightgray}104.6 & \cellcolor{lightgray}99.36 \\
\bottomrule[1.2pt]
\end{tabular}
\label{tab:test_commonvoice_token_ASR}
\end{table}

%% file: tables/7_1_Librispeech_API.tex
\begin{table}[t!]\centering
\small
\renewcommand{\arraystretch}{0.85} 
\setlength{\tabcolsep}{3pt}
\setlength{\abovecaptionskip}{0pt}%
\setlength{\belowcaptionskip}{0pt}%
\caption{[T2] English content protection against \textit{knowledgeable adversary}’s recovery attacks (CRA2).}
\begin{tabular}{llcccc}
\toprule[1.2pt]
\multicolumn{1}{c}{\multirow{2}{*}{\textbf{ASR Model$^\ddagger$}}} & \multicolumn{1}{c}{\multirow{2}{*}{\textbf{Input$^\natural$}}} & \multicolumn{2}{c}{\textbf{Libri. test-clean}} & \multicolumn{2}{c}{\textbf{Libri. test-other}} \\
\multicolumn{1}{c}{} & \multicolumn{1}{c}{} & WER (\%)$\uparrow$ & CER (\%)$\uparrow$ & WER  (\%)$\uparrow$ & CER (\%)$\uparrow$ \\
\midrule[1pt]
\multirow{3}{*}{Wav2Vec2} & Original & 3.15 & 0.88 & 7.68 & 2.72 \\
 & Coded & 3.82 & 1.17 & 11.83 & 4.86 \\
 & \cellcolor{lightgray}\textbf{\sys} & \cellcolor{lightgray}\textbf{101.1} & \cellcolor{lightgray}91.99 & \cellcolor{lightgray}\textbf{101.46} & \cellcolor{lightgray}93.19 \\
\midrule
 
\multirow{3}{*}{Iflytek API} & Original & 8.09 & 4.25 & 13.80 & 6.94 \\
 & Coded & 17.82 & 14.18 & 24.36 & 16.71 \\
 & \cellcolor{lightgray}\textbf{\sys} & \cellcolor{lightgray}98.59 & \cellcolor{lightgray}93.10 & \cellcolor{lightgray}99.54 & \cellcolor{lightgray}93.62 \\
\midrule
 
\multirow{3}{*}{Tecent API} & Original & 4.65 & 3.07 & 8.14 & 4.56 \\
 & Coded & 14.74 & 13.13 & 18.56 & 14.12 \\
 & \cellcolor{lightgray}\textbf{\sys} & \cellcolor{lightgray}99.52 & \cellcolor{lightgray}99.40 & \cellcolor{lightgray}99.68 & \cellcolor{lightgray}99.62 \\
\midrule
 
\multirow{3}{*}{Azure API} & Original & 5.14 & 3.25 & 10.58 & 6.43 \\
 & Coded & 5.68 & 3.51 & 14.56 & 8.95 \\
 & \cellcolor{lightgray}\textbf{\sys} & \cellcolor{lightgray}100.0 & \cellcolor{lightgray}\textbf{99.98} & \cellcolor{lightgray}100.0 & \cellcolor{lightgray}\textbf{100.0} \\
\midrule
\multirow{3}{*}{Amazon API} & Original & 4.98 & 3.24 & 8.56 & 4.80 \\
 & Coded & 15.00 & 13.33 & 19.06 & 14.25 \\
 & \cellcolor{lightgray}\textbf{\sys} & \cellcolor{lightgray}99.86 & \cellcolor{lightgray}95.54 & \cellcolor{lightgray}99.70 & \cellcolor{lightgray}95.07\\ 
\bottomrule[1.2pt]
\end{tabular}
\begin{tablenotes}[flushleft]
    \item[] \vspace{-2pt}\hspace{-2pt}\small 
    (i) $\ddagger$: Here Wav2Vec2 denotes the open-source ASR model~\cite{facebook_wav2vec2}.
    (ii) $\natural$: \textit{Original} means uncompressed audio; \textit{Coded} means the audio go through the OPUS codec processing~\cite{Opus}.
\end{tablenotes}
\label{tab:test_speech_libri_ASR}
\end{table}

%% file: tables/8_commonvoice_API.tex
\begin{table}[t!]\centering
\small
\renewcommand{\arraystretch}{1} 
\setlength{\tabcolsep}{3pt}
\setlength{\abovecaptionskip}{0pt}%
\setlength{\belowcaptionskip}{0pt}%
\caption{[T2] Unseen-language content protection against \textit{knowledgeable adversary}’s recovery attacks (CRA2).}
\begin{tabular}{llccccc}
\toprule[1.2pt]
\multicolumn{1}{c}{\multirow{2}{*}{\textbf{ASR Model$^\ddagger$}}} & \multicolumn{1}{c}{\multirow{2}{*}{\textbf{Input}}} & \multicolumn{5}{c}{\textbf{CVoiceFake WER (\%) $\uparrow$}} \\
\multicolumn{1}{c}{} & \multicolumn{1}{c}{} & English & Chinese & German & French & Italian \\
\midrule[1pt]
\multirow{2}{*}{Wav2Vec2} & Original & 15.69 & 19.03 & 8.93 & 10.24 & 8.38 \\
 & \cellcolor{lightgray}\textbf{\sys} & \cellcolor{lightgray}\textbf{108.47} & \cellcolor{lightgray}90.89 & \cellcolor{lightgray}\textbf{129.49} & \cellcolor{lightgray}\textbf{113.65} & \cellcolor{lightgray}101.51 \\
\midrule
\multirow{2}{*}{Iflytek API} & Original & 18.11 & 7.83 & 18.63 & 25.58 & 31.09 \\
 & \cellcolor{lightgray}\textbf{\sys} & \cellcolor{lightgray}100.39 & \cellcolor{lightgray}97.02 & \cellcolor{lightgray}99.66 & \cellcolor{lightgray}108.8 & \cellcolor{lightgray}\textbf{101.54} \\
\midrule
\multirow{2}{*}{Tencent API} & Original & 11.05 & 7.09 & - & 10.43 & - \\
 & \cellcolor{lightgray}\textbf{\sys} & \cellcolor{lightgray}97.53 & \cellcolor{lightgray}\textbf{100.0} & \cellcolor{lightgray}- & \cellcolor{lightgray}99.66 & \cellcolor{lightgray}- \\
\midrule
\multirow{2}{*}{Azure API} & Original & 10.47 & 10.48 & 14.99 & 20.83 & 8.29 \\
 & \cellcolor{lightgray}\textbf{\sys} & \cellcolor{lightgray}100.0 & \cellcolor{lightgray}100.0 & \cellcolor{lightgray}100.0 & \cellcolor{lightgray}100.29 & \cellcolor{lightgray}99.98 \\
\midrule
\multirow{2}{*}{Amazon API} & Original & 10.45 & 20.44 & 13.60 & 10.99 & 5.93 \\
 & \cellcolor{lightgray}\textbf{\sys} & \cellcolor{lightgray}99.64 & \cellcolor{lightgray}96.06 & \cellcolor{lightgray}99.63 & \cellcolor{lightgray}99.68 & \cellcolor{lightgray}99.55 \\
\bottomrule[1.2pt]
\end{tabular}
\begin{tablenotes}[flushleft]
    \item[] \vspace{-2pt}\hspace{-2pt}\small 
    $\ddagger$: Wav2Vec2 denotes the open-source ASR model~\cite{speechbrain_wav2vec2}; Tecent ASR API does not support German and Italian transcription.
\end{tablenotes}
\label{tab:test_speech_commonvoice_ASR}
\end{table}

%% file: tables/12_reconstructed_STOI.tex
\begin{table}[t!]\centering
\small
\renewcommand{\arraystretch}{1} 
\setlength{\tabcolsep}{1.5pt}
\setlength{\abovecaptionskip}{0pt}%
\setlength{\belowcaptionskip}{0pt}%
\caption{[T2] Speech objective intelligibility (STOI).}
\begin{tabular}{l|cc|ccccc}
\toprule[1.2pt]
\multirow{2}{*}{\textbf{STOI}$^\natural$} & \multicolumn{2}{c|}{\textbf{Librispeech}$\downarrow$} & \multicolumn{5}{c}{\textbf{CVoiceFake}$\downarrow$} \\
& test-clean & test-other & English & Chinese & German & French & Italian \\
\midrule[1pt]
Coded & 0.8698 & 0.8179 & 0.8902 & 0.7844 & 0.7494 & 0.7809 & 0.7326 \\
\rowcolor{lightgray}
\textbf{\sys} & 0.0018 & 0.0015 & 0.0036 & 0.0018 & 0.0106 & 0.0031 & 0.0051\\
\bottomrule[1.2pt]
\end{tabular}
\begin{tablenotes}[flushleft]
    \item[] \vspace{-2pt}\hspace{-2pt}\small 
    (i) $\natural$: The calculation of STOI, which ranges from 0 to 1, is conducted using the original waveform as a reference.
\end{tablenotes}
\label{tab:STOI}
\end{table}

%% file: sections/discussion.tex
\section{Discussion}\label{sec:disucssion}
\color{black}
\hspace{0.4cm}\textbf{Overhead Analysis of \sys.} We evaluate \sys's overhead by comparing its real-time factor (RTF) and floating point operations per second (FLOPs) against established baselines on the identical hardware platform. RTF, defined as $RTF = T_{detect} / T_{audio}$, measures the model's speed in processing audio inputs, where $T_{audio}$ is the duration of the original audio and $T_{detect}$ represents the detection latency. FLOPs reflects the computational complexity of the model---lower FLOPs correspond to lower complexity. As Table~\ref{tab:additional_cost} demonstrates, all methods achieve low RTFs in detecting audio deepfakes. While \sys operates at roughly 2$\sim$3 times the latency of non-privacy-centric methods like AASIST, it significantly outperforms traditional cryptographic methods, which exhibit at least a 100-fold increase in latency over plaintext computations~\cite{chouchane2021privacy}. Regarding FLOPs, despite \sys having slightly higher FLOPs at 62.76T, it remains comparable with other methods. Overall, \sys introduces acceptable additional cost, balancing privacy protection with computational efficiency. We envision that future engineering efforts in model architecture could lead to improvements in overhead.

\input{tables/13v2_additional_cost}

\textbf{Limitation.} (1) For deepfake detection, although \sys demonstrates comparable performance with state-of-the-art detectors, it shares a prevalent limitation in current ML-based detection methods in terms of explainability. (2) For content privacy, though \sys exhibits resilience against various adversaries, as substantiated by our experiments and probabilistic analysis, it is difficult to provide a strong mathematical guarantee since \sys employs a non-cryptographic approach.

\textbf{Probabilistic Perspective Protection.} 
Despite lacking strong mathematical guarantees, \sys protects user content privacy from the probabilistic perspective. Our shuffle layer enhances the CDM that decouples and protects semantic information from exposure to the detection model, forming a dual-layer content privacy protection. Specifically, the shuffle algorithm creates innumerable combinations; for a one-second window of 50 frames, the potential permutations number $50!$ (50 factorial), approximately $3.0414 \times 10^{64}$. Extending this to the entire sequence of acoustic tokens $\mathbf{A}^{b} \in \mathbb{R}^{C\times T_n}$, where $T_n$ is the total number of temporal frames, the complexity expands exponentially as $P_{total} = (50!)^{T_n/50}$. Consequently, the probability of correctly reconstructing a shuffled acoustic token sequence $\mathbf{\overline{A}}$ to its original order $\mathbf{A}$ declines dramatically. For instance, the likelihood of correctly assembling a 4-second audio segment (200 frames) is extremely low, with the probability calculated at $P_{\mathbf{A}} = \frac{1}{(50!)^4} = 1.1687 \times 10^{-258}$. This indicates that our shuffle layer acts as a formidable barrier against content recovery, effectively complementing the protective capabilities of the CDM.

\textbf{Advantages of \sys.} The processing of raw data and the decoupling steps in \sys are lightweight enough to operate on local user devices. However, deepfake detection typically (1) relies on the storage and sharing of confidential audios and (2) needs to be maintained as any large-scale ML model, as in, re-trained and fine-tuned iteratively.

Regarding privacy, if we as a community only develop end-to-end detectors, we remain reliant on raw audios for training, fine-tuning, and validation, and which potentially can be leaked from the trained model. By removing semantic tokens while still on the user's device, the whole detection approach can work on acoustic-only inputs. \sys demonstrates both feasible and operationally effective. This aligns with the concept of ``data minimization’’: if semantic information is not essential for detection, it is prudent to construct a system that obviates its usage. Our talk with mobile vendors has indicated that \sys is recognized as a valuable and attractive feature, enhancing user trust by adding an additional layer of protection to alleviate users’ trust issues towards service/mobile vendors. 

For detection services typically operated by third parties, our method is especially pertinent. It maintains privacy while offering flexible and reliable detection, and can further enable robust decision-making on servers by integrating multiple detection models, which would be computationally heavy if deployed on local user devices. The \sys framework facilitates timely adaptation to deepfake advancements with lower maintenance costs compared to adapting various local devices, thereby safeguarding users from new deepfake risks due to delayed service updates.

\textbf{Dataset for Future Research.} Like the ASVspoof 2019 and 2021 datasets, we plan to release our multilingual CVoiceFake dataset on \cite{safeear_demo} to facilitate research on deepfake detection. The access to CVoiceFake will be granted exclusively to requests adhering to ethical research standards and approved by IRB, for reducing the risk of misusing realistic synthetic audio. Moreover, we advocate for future research to tackle privacy violations in existing applications, establishing privacy-centric intelligent services.
\color{black}



%% file: tables/13v2_additional_cost.tex
\begin{table}[t!]
\centering
\small
\renewcommand{\arraystretch}{1} 
\setlength{\tabcolsep}{10pt}
\setlength{\abovecaptionskip}{0pt}%
\setlength{\belowcaptionskip}{0pt}%
\color{black}
\caption{\label{tab:additional_cost}Additional cost of \sys compared with baseline methods: RTF and FLOPs.}
\begin{threeparttable}
\begin{tabular}{@{}lcc@{}}
\toprule[1.2pt]
\textbf{Method}               & \textbf{RTF} $\downarrow$ & \textbf{FLOPs} $\downarrow$ \\
\midrule[1pt]
AASIST               &  0.0155  &   45.49T   \\
Wav2vec2+Transformer &  0.0111  &   47.05T    \\
SafeEar (Ours)       &  0.0366  &   62.76T    \\
\bottomrule[1.2pt]
\end{tabular}
\vspace{-10pt}
\end{threeparttable}
\end{table}

%% file: sections/related_work.tex
\section{Related Work}
\hspace{0.4cm}\textbf{Defense against Audio Deepfake.}
In the realm of audio deepfake defense, strategies can be divided into three classes: proactive voiceprint anonymization to thwart unauthorized synthesis~\cite{yu2023antifake}, liveness detection leveraging physical properties~\cite{yan2019catcher,li2023towards}, and machine learning (ML)-enabled deepfake detection~\cite{AASIST,RawNet2,Rawformer,LFCC_SE-ResNet34,LCNN-LSTM,ASVspoof2021-baselines}. The research community largely concentrates on ML-based detection systems, given their ease deployment, superior performance and, general applicability. 
To enable accurate ML-based detection systems, prior works extensively explore three aspects: (1) discriminative feature extraction, especially spectral features like MFCC and LFCC~\cite{LFCC_SE-ResNet34,LCNN-LSTM}, and deep learning features like Wav2Vec2~\cite{xie2021siamese}; (2) classification algorithms, \textit{e.g.}, SVM~\cite{alegre2012spoofing}, GMM~\cite{ASVspoof2021-baselines}, CNN~\cite{LFCC_SE-ResNet34}, GNN~\cite{AASIST}, and Transformer~\cite{Rawformer}; (3) generalization methods, \textit{e.g.}, investigating novel loss functions~\cite{chen2020generalization,zhang2021one} and using continual learning strategy~\cite{zeng2023improving} to deal with out-of-domain dataset in real-life scenarios. 
\blue{However, to the best of our knowledge, existing audio deepfake detection systems largely neglect the preservation of speech content privacy. The only exception is a proof-of-concept study employing secure multi-party computation (SMPC), which lacks practicality due to its overly simplistic one-layer architecture and significant latency~\cite{chouchane2021privacy}.}


\textbf{Speech Privacy Preservation.}
Speech privacy preservation efforts are mainly focused on safeguarding speaker voiceprints and speech content. Most existing methods focus on speaker voiceprint protection using signal processing (SP)-based and ML-based anony\-mization methods. SP-based approaches typically involve random perturbations of speech features like MFCC, pitch, and tempo~\cite{patino2020speaker}, or employ uniform transformations~\cite{xiao2023micpro}.
However, these methods often suffer from limited generalizability on out-of-domain speech, leading to compromised quality and unnatural speech output.
ML-based strategies include employing TTS/VC systems for voiceprint alteration~\cite{justin2015speaker} or mapping speeches to an anonymized and average voiceprint style~\cite{bahmaninezhad2018convolutional,wang2023vsmask}. Additionally, adversarial examples (AE) have proven effective in misguiding traditional speaker verification systems~\cite{deng2023v,ze2023ultrabd,li2023enrollment}. Yet, none of these approaches adequately protect speech content, particularly from human auditory analysis. While Preech~\cite{ahmed2020preech} considers protecting partial content privacy by using an extra local ASR model to substitute sensitive words, it may fail to identify sensitive content in noisy environments. Moreover, its TTS/VC-based dummy word injection strategy results in an unnatural blend of genuine and synthesized speech segments, which could hinder deepfake detection efforts.


\textbf{Our Approach.} \sys fills a critical void in the realm of privacy-preserving audio deepfake detection. It ensures the confidentiality of content by decoupling semantic and acoustic tokens, subsequently shuffling the latter to provide a dual layer of protection. Employing solely shuffled acoustic tokens, \sys effectively detects deepfakes through the implementation of real-world codec augmentation strategies.

%% file: sections/conclusion.tex
\section{Conclusion}
In this paper, we investigate the intersections of deepfake detection and privacy preservation. Specifically, we introduce \sys, a novel framework that realizes effective audio deepfake detection while preserving speech content privacy. The key idea of \sys lies in decoupling speech information into discrete semantic and acoustic tokens, and further adopting the shuffling method to form a dual protection against machine and human analysis. \blue{We enhance the acoustic-only deepfake detector with optimal MHSA's heads and real-world codec augmentation} to enable effective deepfake detection only based on the shuffled acoustic tokens. The efficacy of \sys is validated through extensive testing on our established benchmark, achieving an EER of 2.02\%. It can also protect multilingual content from a series of \textit{content recovery adversaries}, as evidenced by the 93.9\% WERs alongside our user study.

%% file: sections/appendix.tex
\appendix
\section{Audio Codec}\label{ssec:audio_codec}
Audio codecs are widely used in the real-time communication tools and media softwares, which compress and decompress audio data from a live stream media (such as radio) or an already stored data file. The purpose of using an audio codec is to effectively reduce the size of an audio file without affecting the quality of the sound. There are two categories of audio codecs:

\textbf{Traditional codecs:} traditional digital signal processing (DSP) codecs, such as MP3~\cite{MP3}, Opus~\cite{Opus}, AAC~\cite{AAC}, G.722~\cite{G.722} and Ogg Vorbis~\cite{Ogg}, are integral in telecommunications, streaming, and broadcasting. These codecs utilize mathematical techniques, \textit{e.g.}, subband modulation~\cite{G.722}, psychoacoustic modeling~\cite{MP3,AAC}, and transform coding~\cite{MDCT}, to remove audio components that are less likely to be perceived by the human ear to achieve compression. Although traditional DSP codecs remain widely used due to their compatibility and ease of use, they face limitations, such as suboptimal compression efficiency and compromised quality at low bitrates.

\textbf{Neural Codecs:} compared with traditional codecs, neural audio codecs, such as Encodec~\cite{Encodec} and SoundStream~\cite{Soundstream}, offering multi-aspect advantages, including audio type-agnostic and real-time operation that can effectively encode and decode various sound types, \textit{e.g.}, clean, noisy and reverberant speech, music and environmental sounds, with no additional latency. The most significant feature is their state-of-the-art sound quality over a broad range of bitrates. Traditional codecs introduce coding artifacts at poor network connectivity (\textit{i.e.}, low bitrates), while neural codecs~\cite{Encodec} can operate even at low bitrates from 1.5kbps to 24kbps, with a negligible quality loss. This attributes to its training with structured multi-layer residual vector quantizers (RVQs). 

Pioneered by VQ-VAE \cite{van2017neural}, the RVQ concept for discrete speech representation has inspired a new paradigm in codec-based audio generation, exemplified by models like AudioLM \cite{borsos2023audiolm}, VALL-E \cite{wang2023neural}, HAM-TTS \cite{hamtts}, and USLM \cite{zhang2023speechtokenizer}. The codec efficiently encodes speech into fixed-dimension tokens for further application in TTS and VC domains.
We make the first attempt to design neural codec-based discrete tokens for deepfake detection, where our distinctive contribution lies in the design of a decoupling strategy for semantic and acoustic tokens within RVQs. This strategy is pivotal for enabling \sys to execute privacy-preserving detection without semantic information leakage.

\section{Speech Content Recognition}\label{appendix:ASR}
An automatic speech recognition (ASR) system aims to transcribe the speech contents from audio samples. It functions by first segmenting the audio input into discrete frames and carefully extracting speech features; then employs probabilistic models to assign likelihoods to each frame's features that designate potential correspondences with specific phonemes or words. This vital process decodes the feature representation flow of speech inputs through to the output of textual transcription. As for the forms of speech features, they have evolved through significant shifts, pivoting from mathematically crafted Filter Bank (FBank), Constant-Q, Linear-frequency, and Mel-frequency cepstral coefficients (CQCC, LFCC, and MFCC), using neural encoders to learn suitable speech representations, as well as employing self-supervised models like Wav2Vec2 and Hubert. There has also been a marked enhancement in the probabilistic models used in ASR systems, evolving from DNNs~\cite{hinton2012deep}, to long-short term memory networks (LSTM)~\cite{graves2013hybrid}, and on to Conformers~\cite{gulati2020conformer}. This progression has substantially strengthened the model's capability to represent the probabilistic transitions between phonemes (\textit{i.e.}, from features to text).

\input{sections/loss_function}

\section{Tandem Detection Cost Function (t-DCF)}\label{appendix:t-DCF}
The tandem Detection Cost Function (t-DCF) provides a metric for assessing the efficiency of deepfake countermeasures under varied conditions, especially in the realm of speaker verification systems. It effectively combines the impact of misses (i.e., failing to detect a genuine attempt) and false alarms (i.e., incorrectly flagging a deepfake attempt as genuine) into a single cost figure. The t-DCF is calculated using the following equation:

\begin{equation}
\text{t-DCF} = C_{miss} \cdot P_{miss}^{\text{cm}} \cdot P_{target} + C_{fa} \cdot P_{fa}^{\text{cm}} \cdot (1 - P_{target})
\end{equation}

In this equation, $C_{miss}$ and $C_{fa}$ represent the cost of misses and false alarms, respectively. $P_{miss}^{\text{cm}}$ denotes the miss rate of the countermeasure, $P_{fa}^{\text{cm}}$ signifies the false alarm rate, and $P_{target}$ represents the a priori likelihood of encountering a genuine target trial in a speaker verification scenario. This cost function reflects the weighted importance of error rates in the decision-making process of a system, offering a nuanced view of the practical performance of countermeasure mechanisms against deepfake attempts in speaker authentication.


%% file: sections/loss_function.tex
\section{Loss Functions of Codec-based Decoupling Model}\label{sec:CDM_loss}
To better decouple the semantic and acoustic information of the input audio, we introduce multiple loss functions, including distillation loss, reconstruction loss, perceptual loss derived from the discriminator, and RVQ commitment loss. 

The purpose of distillation loss is to extract semantic information from the audio. And then we aim to modify the first quantizer (VQ1) to capture the semantic information from speech, serving a content-centric role. Specifically, we introduce a knowledge distillation approach, \textit{i.e.}, employing the well-established HuBERT~\cite{hsu2021hubert} as our semantic teacher of VQ1. Since HuBERT can well represent given speech as semantic-only features~\cite{mohamed2022self}, we employ the average representation across all HuBERT layers as the semantic supervision signal that encourages the semantic student VQ1 to learn a very close content representation via:
\begin{equation}
    \mathcal{L}_{distill}=\frac{1}{T_n}\sum_{t=1}^{T_n}\log\sigma(\cos{(\mathbf{W}\cdot\mathbf{S}_t,\mathbf{H}_t)})
\end{equation}
where $\mathbf{S}_t$ and $\mathbf{H}_t$ respectively denote the $t^{th}$ quantized output, \textit{i.e.},  $t^{th}$ token frame of the VQ1 and the HuBERT. $\cos(\cdot)$ is cosine similarity. $\sigma(\cdot)$ denotes sigmoid activation. $\mathbf{W}$ is the projection matrix.

The reconstruction loss consists of two parts: the time domain and the frequency domain. In the time domain, the aim is to minimize the L1 distance between the original audio $X$ and the reconstructed audio $\hat{X}$. In the frequency domain, on the other hand, we take a more nuanced approach that involves a linear combination of L1 and L2 losses on the mel-spectrogram at different time scales. This approach aims to capture and minimize the difference in frequency characteristics between the target and generated audio. Formally, the reconstruction loss can be expressed as:
\begin{equation}
    \mathcal{L}_{{\text{rec}}}=\sum_{i \in e}(\left\|\mathcal{M}_{i}(X)-\mathcal{M}_{i}(\hat{X})\right\|_{1}+\left\|\mathcal{M}_{i}(X)-\mathcal{M}_{i}(\hat{X})\right\|_{2})+||X-\hat{X}||_1,
\end{equation}
where $\mathcal{M}_i(\cdot)$ denotes the mel-spectrogram using STFT with different window sizes $2^i$ and hop sizes $2^i//4$, $i\in [5,11]$.

We introduce the adversarial loss to learn the features of real audio more efficiently and thus generate high-quality audio under different discriminator evaluations. This strategy not only improves the realism of the generated audio but also enhances the robustness of the model in complex audio generation tasks. Specifically, we compute the losses of multiple discriminators and perform time averaging to obtain a combined adversarial loss value. Formally, this adversarial loss can be expressed as:
\begin{align}
     \mathcal{L}_{{\text{G}}}&=\frac{1}{K} \sum_{k=1}^{K} \max \left(1-D_{k}(\hat{X}), 0\right), \\
    \mathcal{L}_{\text{D}}&=\frac{1}{K} \sum_{k=1}^{K} \max \left(1-D_{k}(X), 0\right)+\max \left(1+D_{k}(\hat{X}), 0\right),
\end{align}
where $K$ denotes the number of discriminators $D_k(\cdot)$. In addition, we also add a relative feature matching loss \cite{defossez2022high} to the generator:
\begin{align}
\mathcal{L} _{\text {feat }}(X, \hat{X})=\frac{1}{K L} \sum_{k=1}^{K} \sum_{l=1}^{L} \frac{\left\|D_{k}^{l}(X)-D_{k}^{l}(\hat{X})\right\|_{1}}{\operatorname{mean}\left(\left\|D_{k}^{l}(X)\right\|_{1}\right)},
\end{align}
where $L$ denotes the number of layers in discriminators.

For the RVQ, we introduce a computation of the commitment loss $\mathcal{L}_c$ between the pre-quantized and quantized values. Note that the quantized values do not compute the gradient. This training objective can be formulated as follows:
\begin{equation}
    \mathcal{L}_c=\sum_{N_q}^{i=1}\left \| \mathbf{z}_i - q(\mathbf{z}_i) \right \|^2_{2}
\end{equation}

In summary, the DCM model's generator part is trained to optimize the following loss:
\begin{equation}
    \mathcal{L}_\text{gen}= \lambda_{\text{d}} \mathcal{L}_{\text{distill}} + \lambda_{\text{r}} \mathcal{L}_{\text{rec}} + \lambda_{\text{G}} \mathcal{L}_{\text{G}} + \lambda_{\text{f}} \mathcal{L}_{\text{feat}} + \lambda_{\text{c}} \mathcal{L}_{\text{c}}
\end{equation}
where we set coefficients similar to HiFiGAN~\cite{kong2020hifi}, with specific values $\lambda_{\text{d}}=1, \lambda_{\text{r}}=1, \lambda_{\text{G}}=3, \lambda_{\text{f}}=3, \lambda_{\text{c}}=1$.